\documentclass[10pt,a4paper,onecolumn]{article}
\usepackage[T1]{fontenc}
\usepackage{amssymb}
\usepackage[utf8]{inputenc}
\usepackage{setspace}
\usepackage[font=footnotesize]{caption}
\usepackage[biblabel]{cite}
\usepackage{graphicx}
\usepackage{float}
\usepackage{lineno}
\usepackage{xcolor}
\usepackage{multicol}
\setpagewiselinenumbers
\let\oldbibliography\thebibliography
\renewcommand{\thebibliography}[1]{%
  \oldbibliography{#1}%
  \setlength{\itemsep}{0pt}%
}
\graphicspath{{Figs/}}

\begin{document}
\pagenumbering{arabic}

\begin{center}
{\Large{\bf {\Large{\bf Transition metal dichalcogenide dimer nano-antennas with ultra-small gaps}}}}
\vskip1.0\baselineskip{Panaiot G. Zotev$^{a*}$, Yue Wang$^{b**}$, Luca Sortino$^{a,c}$, Toby Severs Millard$^a$, Nic Mullin$^a$, Donato Conteduca$^b$, Mostafa Shagar$^a$, Armando Genco$^a$, Jamie K. Hobbs $^a$, Thomas F. Krauss$^b$, Alexander I. Tartakovskii$^{a***}$}
\vskip0.5\baselineskip\footnotesize{\em$^a$Department of Physics and Astronomy, University of Sheffield, Sheffield S3 7RH, UK\\$^b$Department of Physics, University of York, York, YO10 5DD, UK\\$^c$Chair in Hybrid Nanosystems, Nanoinstitute Munich, Faculty of Physics, Ludwig-Maximilians-Universit\"{a}t, M\"{u}nchen, 80539, Munich, Germany\\}
$^{*}$p.zotev@sheffield.ac.uk \quad $^{**}$yue.wang@york.ac.uk \quad $^{***}$a.tartakovskii@sheffield.ac.uk

\end{center}
\vskip1.0\baselineskip

\begin{multicols}{2}

\textbf{Transition metal dichalcogenides have emerged as promising materials for nano-photonic resonators due to their large refractive index, low absorption within a large portion of the visible spectrum and compatibility with a wide range of substrates. Here we use these properties to fabricate WS$_2$ double-pillar nano-antennas in a variety of geometries enabled by the anisotropy in the crystal structure. Using dark field spectroscopy, we reveal multiple Mie resonances, to which we couple WSe$_2$ monolayer photoluminescence and achieve Purcell enhancement and an increased fluorescence by factors up to 240. We introduce post-fabrication atomic force microscope repositioning and rotation of dimer nano-antennas, achieving gaps as small as 10$\pm$5 nm, opening the possibility to a host of potential applications including strong Purcell enhancement of single photon emitters and optical trapping, which we study in simulations. Our findings highlight the advantages of using transition metal dichalcogenides for nano-photonics by exploring new applications enabled by their unique properties.}\\

Transition metal dichalcogenides (TMDs) have drawn large scientific and technological interest in the past decade since the discovery of a direct band gap in monolayers due to quantum confinement effects \cite{Mak2010}, which in conjunction with reduced dielectric screening leads to strongly bound excitons \cite{Wang2018}. These layered materials found their way to research involving integration with nano-photonic structures such as plasmonic and dielectric cavities to realize both weak and strong coupling \cite{Xia2014,Dufferwiel2018,Zhang2018,Krasnok2018,Sortino2019}, low-threshold lasing \cite{Wu2015}, Purcell \cite{Cai2018a,Luo2018} and quantum efficiency enhancement \cite{Sortino2021} of single photon emitters (SPEs) as well as coupling to collective resonances found in periodic structures \cite{Bernhardt2020,Kravtsov2020}. In these studies, the use of TMDs was limited to single and few-layer samples focusing on coupling emitted light from 2D semiconductors to resonances and cavity modes in different material systems \cite{Mak2016a}.

Alternatively, the fabrication of a photonic resonator from a layered material similar to TMDs was first achieved in hexagonal boron nitride (hBN). Recent reports utilized electron beam induced and reactive ion etching of hBN to fabricate suspended one- and two-dimensional photonic crystal cavities as well as ring resonators, circular Bragg gratings and waveguides \cite{Kim2018a,Froch2019}. Moreover, hBN microcavity-like structures have been shown to control the spontaneous emission rate of excitons in MoS$_2$ monolayers \cite{Fang2019}. Micro-rotator structures, twisted with an atomic force microscope (AFM) cantilever tip, have also been shown to facilitate the control of second harmonic generation (SHG) enhancement in hBN \cite{Yao2021} as well as the properties of novel electronic devices \cite{Ribeiro-Palau2018}. Photonic resonators fabricated in TMDs, however, have only recently been reported even though these materials offer a number of advantages. The refractive index of WS$_2$ in the visible range (n>4) \cite{Verre2019} is higher than that of hBN (n $\approx$ 2.1) \cite{Rah2019} or other high-index dielectrics traditionally used to fabricate nano-photonic resonators such as gallium phosphide (n $\approx$ 3.5)) \cite{Cambiasso2017} or silicon (n $\approx$ 3.8)) \cite{Cambiasso2017,Bakker2015a,Cambiasso2018,Xu2018}. TMDs also often maintain a sizable transparency window in the visible \cite{Li2014a} and offer advantages due to their layered nature such as large optical anisotropy \cite{Ermolaev2021} and adhesion to a large variety of substrates owing to their van-der-Waals attractive forces \cite{Frisenda2018}. These properties offer the possibility of producing a highly contrasting refractive index boundary by deposition of TMD crystals onto low refractive index materials such as SiO$_{2}$ \cite{Benameur2011} thereby providing a straightforward route to highly confined optical resonances.

Recent reports of TMD photonic structures have demonstrated strong coupling using WS$_{2}$ photonic crystals \cite{Zhang2020}, gratings \cite{Zhang2020a}, nano-antenna resonators \cite{Verre2019} as well as TMD bulk flakes \cite{Munkhbat2019}. Waveguiding or quasi-waveguiding has also been achieved in monolayer WS$_{2}$ photonic crystals \cite{Zhang2019} and bulk TMD flakes \cite{Ermolaev2021}. TMD nano-disk Mie resonators, hosting non-radiative anapole modes, have also been fabricated to show second and third harmonic generation enhancement \cite{Busschaert2020} and Raman scattering enhancement \cite{Green2020}. Numerical studies have explored the possibility of entire optical circuits in TMDs including rib waveguides, photonic crystal cavities and electro-optic modulators \cite{Ling2021}. Further theoretical reports open the possibility for realization of MoS$_{2}$ nano-resonator modes \cite{Muhammad2021} as well as WS$_{2}$ nano-antenna metasurface resonances \cite{Ahmed2020} characterized as bound states in continuum.

Photonic structures with ultra-small gaps and a double-vertex geometry are highly desirable due to the possibility of strong confinement of electric and magnetic fields due to the boundary conditions on the normal and parallel components of the electric field at a sharp refractive index contrasting boundary\cite{Choi2017}. The high fields are a prerequisite for large radiative rate enhancements of emitters as evidenced from tip cavity structures in photonic crystal nano-beam cavities \cite{Choi2017} and plasmonic bowtie antennas \cite{Kinkhabwala2009}. Incidentally, such large electric field intensities can also lead to stable optical trapping and therefore precise positioning of nano-particles such as quantum dots or polystyrene beads, which closely resemble the size and refractive index of large proteins. This is due to an attractive Lorentz force in the direction of an electromagnetic hotspot under optical excitation, which is dependent on the particle size, refractive index, the input pump power and the energy confinement provided by the photonic environment\cite{Conteduca2017}. Previous reports of nano-antenna optical trapping utilize plasmonic resonators \cite{Wang2011}, however, they suffer from large changes in temperature leading to loss in stability as well as quenching of emission due to increased optical absorption processes \cite{Xu2018,Jensen2016}. Alternatively, dielectric nano-resonators, such as silicon dimer nano-antennas \cite{Xu2018,Xu2019}, can be advantageous for optical trapping in different applications including biological nano-particles at risk of degrading due to heating effects as well as quantum dot positioning without emission quenching. The large field confinement in closely spaced double-vertex structures, which is advantageous for Purcell enhancement as well as optical trapping, may be achieved in WS$_2$ by using the etching anisotropy of the crystallographic axes \cite{Munkhbat2020} and the weak van-der-Waals forces.

In this work, we pattern nano-antenna structures into thin WS$_{2}$ crystals. We exfoliate 25 nm - 500 nm thick flakes of WS$_{2}$ onto a SiO$_{2}$ substrate and utilize established nano-fabrication techniques such as electron beam lithography (EBL) and reactive ion etching (RIE) to define sub-micron nano-antennas with nanometer scale gaps. We observe that WS$_{2}$ can be selectively fabricated in circular, square or novel hexagonal geometries with potentially atomically sharp edges and vertices depending on the etching recipe used. Dark field spectroscopy of single (monomer) and double (dimer) nano-pillar resonators reveals geometric Mie resonances, which we compare with finite-difference time-domain (FDTD) simulations. We transfer a monolayer of WSe$_2$ onto an array of fabricated dimer nano-antennas and observe photoluminescence enhancement factors of more than 240 on the structures when compared to emission from regions on flat SiO$_2$. We also observe polarization dependent PL emission aligned with the dimer axis and lifetime shortening by a factor of nearly 2 confirming the coupling of the WSe$_2$ monolayer emission to the photonic resonances of the nano-antennas and yielding a Purcell factor lower bound of 1.85. Subsequently, we utilize contact mode atomic force microscopy (AFM) as a post-fabrication step to reposition the constituent nano-pillars of dimer nano-antennas achieving gaps of 10$\pm$5 nm. We further numerically study the viability of utilizing dimer nano-antennas for the enhancement of single photon emission rates. We simulate the electric field confinement as well as the Purcell factor for an SPE positioned at the hotspots of the dimer nano-antenna mode. These simulations yield electric field intensity enhancement of >10$^3$ compared to vacuum and Purcell factors of >150 for hexagonal and square geometries. We subsequently numerically explore the prospect of using WS$_2$ dimer nano-antennas with ultra-small gaps in optical trapping. For the smallest experimentally achieved dimer gap, we calculate attractive forces towards the electric field hotspots of >350 fN for colloidal quantum dots (QDs) and >70 fN for polystyrene beads (PBs) which closely emulate large proteins. Our experimental and numerical studies of TMD material photonic resonators opens possibilities for novel methods of radiative rate enhancement and optical trapping which may lead to a scalable route of fabricating Purcell enhanced SPEs for a variety of applications including quantum computing and communication.\\

\large
\noindent\textbf{Results}\\
\normalsize

\begin{figure}[h!]
	\centering
  \includegraphics[width=\linewidth]{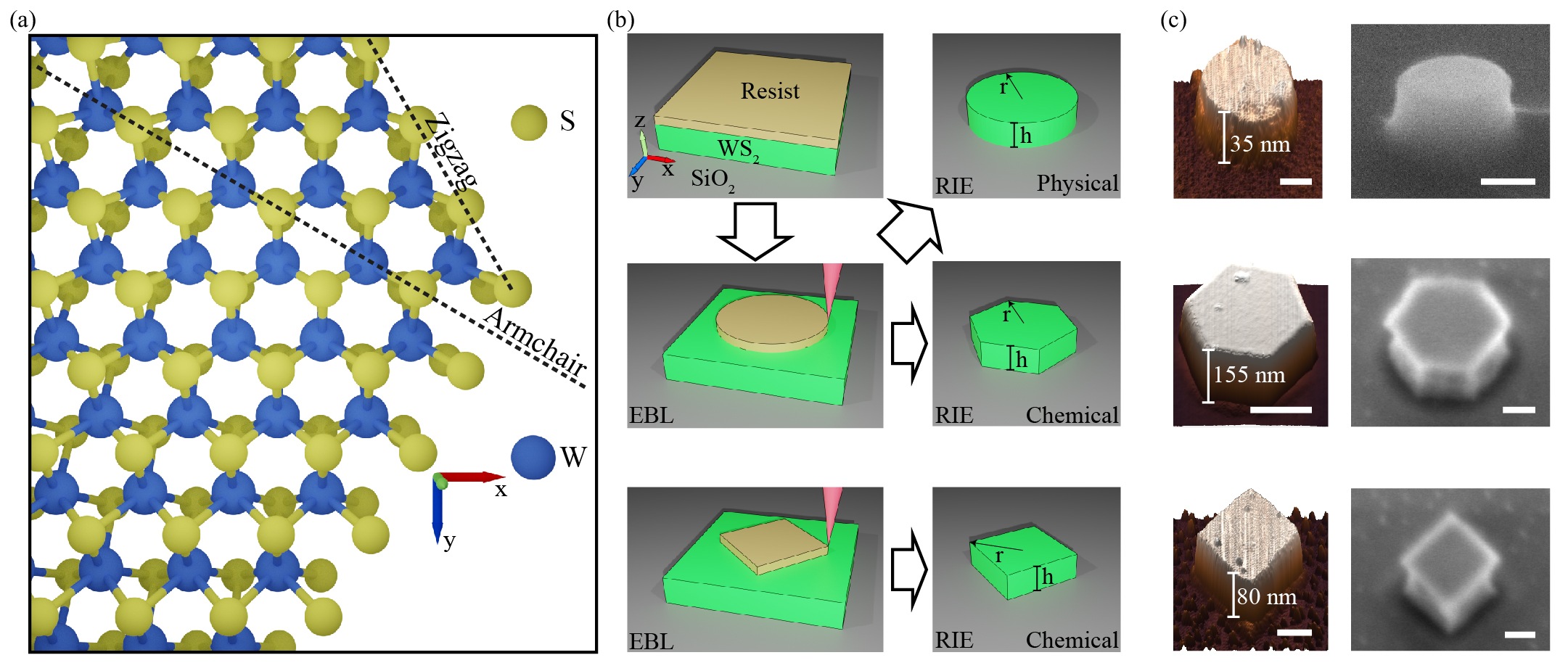}
  \caption{\textbf{Fabrication procedure for WS$_{2}$ nano-antennas.}
	\textbf{(a)} Representation of a top view of a single layer of WS$_{2}$ found at the top of a hexagonal nano-antenna at the position of the vertex. The zigzag axis defines the edges of the structure due to its higher stability as shown by the two bonds on outlying sulfur atoms when compared to the armchair axis with its single bond for outlying sulfur atoms. 
	\textbf{(b)} Fabrication steps and their order following the black outline arrows. The first step includes spinning of resist onto a WS$_{2}$ flake. The second step is patterning and development after electron beam lithography into a circular or square geometry. The final step is reactive ion etching using either a more physical or chemical etching recipe. The height (h) is defined by the thickness of the original flake. The radius is defined as the distance from the center of the structure to an outside vertex for hexagonal and square geometries. 
	\textbf{(c)} AFM and SEM images of fabricated nano-antennas. The left column shows 3D representations of AFM scans of circular, hexagonal and square nano-antennas in the top, middle and bottom row respectively. Scale bars in AFM scans = 200 nm. The right column shows SEM images of similar structures taken at a 60$^\circ$ tilt of the sample. Scale bars in SEM images = 100 nm.}
  \label{F1}
 \end{figure}

\noindent\textbf{Fabrication of nano-antennas} WS$_{2}$ consists of covalently bonded monolayers (Fig. \ref{F1}(a)) with a hexagonal crystal structure held together by van-der-Waals forces in bulk crystals. We mechanically exfoliated WS$_{2}$ flakes onto a 290 nm SiO$_{2}$ on silicon substrate with thicknesses ranging from 25 to 500 nm. Fig. \ref{F1}(b) shows a schematic representation of the subsequent fabrication process. We spun a positive resist onto the sample and patterned circular disks and squares with varying radii using EBL. After development, we transferred the pattern into the WS$_{2}$ crystals with RIE using two different recipes. Etching was terminated once the etch depth (estimated from the etching rate and time) matched the thickness of the exfoliated crystal, defining the nano-antenna height (see Methods). An anisotropic etch using a mixture of CHF$_{3}$ and SF$_{6}$ gases along with a high DC bias and low chamber pressure yielded circular nano-pillars with vertical sidewalls, which resulted from the physical etching of the resist pattern into the WS$_{2}$. Examples of completed structures are shown in the AFM and scanning electron microscopy (SEM) images in the upper row of Fig. \ref{F1}(c). 

Upon substitution of the CHF$_{3}$ with additional SF$_{6}$ gas, a reduction of the DC bias and an increase in the chamber pressure, the resulting nano-antennas exhibited a hexagonal geometry with a radius defined from the center of the structure to one of the vertices at the edge, as shown in the middle right panel of Fig. \ref{F1}(b). This definition corresponds to the radius of the previously circular resist pattern. The physical etching mechanism was suppressed and the increased proportion of reactive fluorine radicals ensured a dominant chemical etch, which preferentially removed the WS$_{2}$ crystal in the armchair crystal axis leading to zigzag terminated sidewalls at 120$^{\circ}$ angles to each other following the crystal symmetry. This agrees with DFT results, which also predict that zigzag edges are more stable \cite{Li2008,Xiao2016}. The AFM and SEM images in the middle row of Fig. \ref{F1}(c) display an example of hexagonal nano-antennas. 

The final geometry we achieved was that of a square which resulted from a combination of resist patterning and chemical etching of the WS$_{2}$. We defined a square resist pattern with sides oriented parallel to the zigzag axis of the crystal. The subsequent chemical etching similarly removed the WS$_{2}$ in the armchair crystal axes ultimately leading to 90$^{\circ}$ angles describing a square shaped nano-antenna, examples of which are displayed in the lower panels of Fig. \ref{F1}(c). The hexagonal geometry and, in part, the square nano-antenna geometry are formed due to the relative stability of the zigzag axis and can therefore lead to atomically sharp vertices.\\

\noindent\textbf{Photonic resonances of WS$_{2}$ nano-antennas} We studied the fabricated structures using dark field spectroscopy and compared the experimental results to FDTD simulations, which yielded close agreement. We identified an electric dipole resonance with small contributions from higher order modes as well as anapole and higher order anapole modes (see Supplementary Note 1). As expected from Mie theory, the resonances forming in the studied nano-antennas red-shifted with increasing radius and blue-shifted with decreasing height. The ability to change the geometry of nano-antennas through a choice of etching recipe provides an additional, more precise tuning mechanism for the observed resonances (see Supplementary Note 2). 

\begin{figure}[h!]
	\centering
  \includegraphics[width=\linewidth]{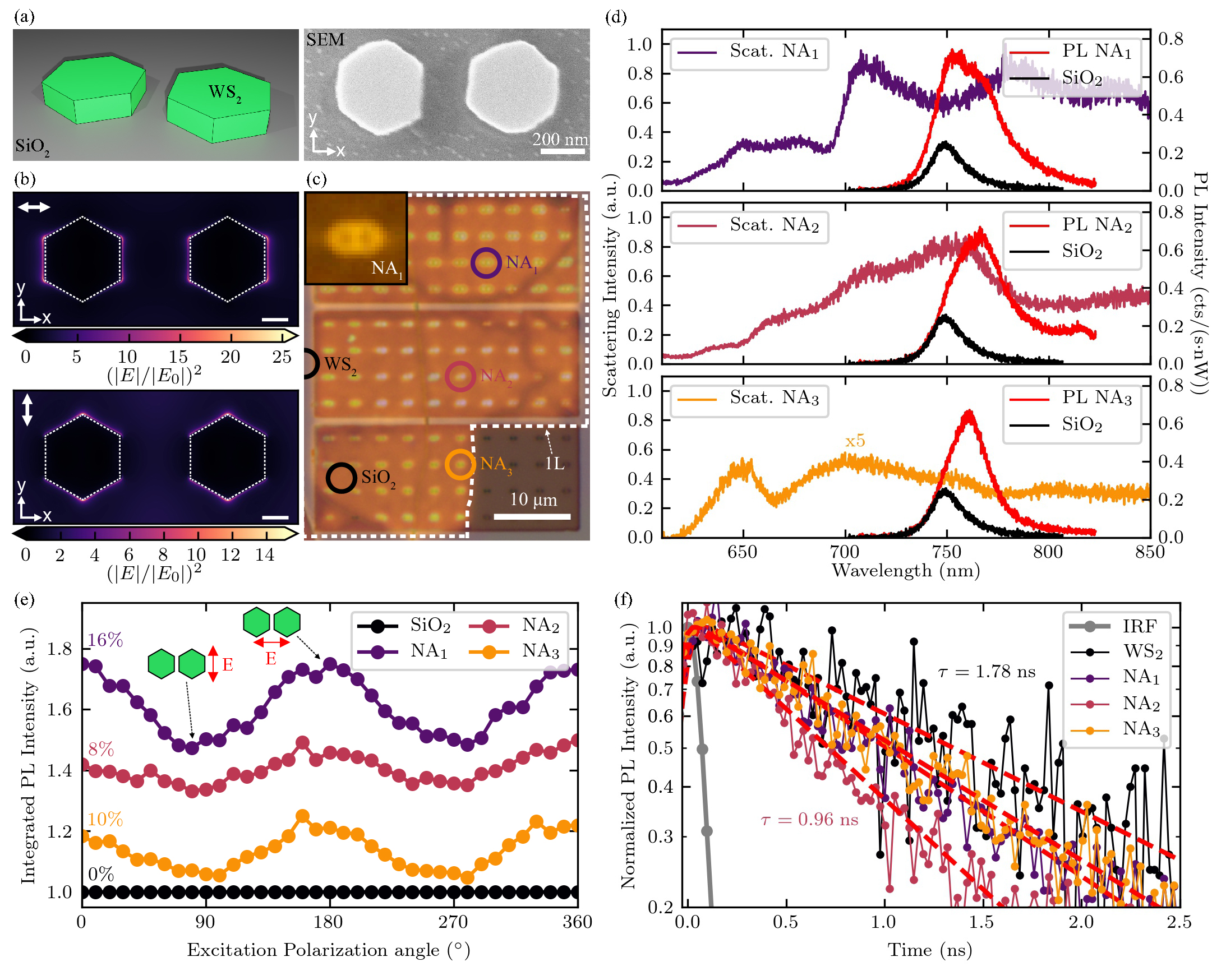}
  \caption{\textbf{PL emission enhancement from monolayer WSe$_2$ on WS$_2$ hexagonal dimer nano-antennas.}
	\textbf{(a)} Left panel: Schematic illustration of a hexagonal dimer nano-antenna. Right panel: SEM image of the top view of a fabricated hexagonal dimer nano-antenna.
	\textbf{(b)} Top view spatial distributions of the simulated electric field intensity 0.5 nm from the top surface of a WS$_2$ dimer nano-antenna with a radius of 165 nm, a gap of 200 nm and a height of 135 nm. Incident plane wave polarization is parallel to the double white arrow in the top left corner of each panel. Hotspots of high electric field intensity form on the edges towards the gap and away from it in the upper panel (X-pol mode). Hotspots form at top and bottom edges of the nano-structure in the lower panel (Y-pol mode) yielding less confinement for this plane wave polarization. Axes (x,y) indicate cross sectional surface similar to that in right panel of \textbf{(a)}. Dashed white outlines represent the physical edges of the structures. Scale bars = 100 nm.
	\textbf{(c)} Photoluminescence microscope image overlaid on a bright field microscope image showing a map of the studied sample including the monolayer region. Bright emission is observed at nano-antenna sites. Inset: Magnified PL image of NA$_1$ showing that bright emission regions follow the outline of the nano-antenna geometry.
	\textbf{(d)} Dark field scattering spectra for three exemplary nano-antennas shown with colored circles in \textbf{(c)}. Photoluminescence spectra from the WSe$_2$ monolayer on the nano-antenna sites (red) as well as from a flat region on SiO$_2$ (black) are displayed as well. The PL intensity on all nano-antenna sites is enhanced compared to that from flat SiO$_2$. The overlap between the scattering resonances and the monolayer PL spectrum from the nano-antenna sites suggests enhanced emission.
	\textbf{(e)} Excitation polarization dependent integrated PL measured at the three nano-antenna sites. Degrees of linear polarization are shown on the left side of each curve. All nano-antennas exhibit a polarization dependence, which is aligned to the dimer axis suggesting coupling between the monolayer emission and the nano-antenna resonances. The intensity is normalized to that from a flat monolayer on SiO$_2$ and vertically offset for illustration purposes.
	\textbf{(f)} Photoluminescence decay measured from the monolayer at each nano-antenna site and on bulk WS$_2$ crystal. The measured lifetimes are as follows: WS$_2$: $1.78\pm0.08$ ns, NA$_1$: $1.34\pm0.02$ ns, NA$_2$: $0.96\pm0.01$ ns, NA$_3$: $1.45\pm0.02$ ns. Reduced lifetimes on nano-antennas suggests Purcell enhancement of emission.}
  \label{F2}
\end{figure}

Next we considered more complex architectures by placing a pair of the novel hexagonal nano-pillars in close proximity to form a dimer nano-antenna shown schematically in the left panel of Fig. \ref{F2}(a). An SEM image of a fabricated structure is also displayed in the right panel of Fig. \ref{F2}(a). Dark field spectra of dimer nano-antennas also exhibit scattering Mie resonances as well as anapole modes which redshift with increasing size (see Supplementary Note 3). However, when two single nano-pillars are placed in close proximity to form a dimer, their photonic resonances hybridize forming two cross-polarized modes with an energy splitting (see Supplementary Note 4). One of these modes is excited with a polarization parallel to the axis connecting the midpoints of the single nano-pillars (dimer axis), which we name the X-pol mode, while the other is excited perpendicularly, here named the Y-pol mode. An intriguing result from this hybridization is the increased electric field intensity surrounding the side-walls of the nano-antennas. For the X-polarized mode, high electric field intensity hotspots form in the gap separating the two nano-pillars as shown in the upper panel of Fig. \ref{F2}(b). For the Y-polarized mode, these hotspots form at the top and bottom wall of each nano-pillar as shown in the lower panel of Fig. \ref{F2}(b). This confinement of the electric field intensity is expected to induce a large density of optical states suggesting that these nano-antennas are advantageous for Purcell enhancement of emission. \\

\noindent\textbf{WSe$_{2}$ monolayer photoluminescence enhancement} In order to investigate whether these nano-structures can be used for such an application, we transferred a monolayer of WSe$_2$ onto an array of WS$_2$ dimer nano-antennas with a varying radius and gap distance using an all-dry technique (see Methods). A photoluminescence (PL) image of the completed sample is shown in Fig. \ref{F2}(c). The brighter emission surrounding the nano-antenna sites is a first indication of the enhanced PL emitted by the monolayer due to an interaction with the dimer nano-antennas. The shape of this bright emission follows the outer walls of the nano-antennas, as seen in the inset of Fig. \ref{F2}(c), similar to the calculated higher electric field intensity regions shown in Fig. \ref{F2}(b).

In order to experimentally evaluate the photonic response of the structures, we subsequently measured the dark field spectra of three nano-antennas with a monolayer of WSe$_2$ on top. These are displayed in Fig. \ref{F2}(d) together with PL emission from the monolayer measured at each nano-antenna site. These structures have a height of 135 nm, a gap of 150 nm as well as a range of radii (NA$_1$: r = 235 nm, NA$_2$: r = 185 nm, NA$_3$: r = 120 nm), representing an exemplary set of the measured nano-antennas from our sample, which included 86 dimers with nominal geometries corresponding to one of the three presented here. The overlap between the scattering Mie resonances and the PL spectrum suggests that all of the nano-structures may induce some emission enhancement; however, the strongest effect is expected from dimer NA$_2$.

We subsequently carried out detailed room temperature photoluminescence measurements in a micro-PL setup in order to study the enhanced emission from the WSe$_2$ monolayer in more detail. The excitation source, a pulsed laser (80 MHz) at 638 nm, was chosen to be below the WS$_2$ absorption edge so that it would only be absorbed in the WSe$_2$ monolayer and not in the nano-antennas. The spectra recorded at the position of the three dimer nano-antennas are displayed in Fig. \ref{F2}(d) (red). These are compared to a PL spectrum measured from a flat portion of monolayer on SiO$_2$ shown in black. The redshift of the PL spectrum observed here is due to strain in the monolayer as it conforms to the nano-antenna geometry \cite{Sortino2020}. This suggests that the WS$_2$ dimer nano-antenna platform is ideal to study strain effects in monolayer TMDs similar to dimers fabricated from other dielectrics \cite{Sortino2020,Sortino2021}. The luminescence intensity from the monolayer at the nano-antenna positions is 3.5 to 4 times brighter. As the excitation spot is much larger than the nano-structures, we defined an experimental enhancement factor $\left\langle EF \right\rangle$, similar to reference \cite{Sortino2019}, in order to estimate the enhanced PL intensity:

\begin{equation}
\left\langle EF \right\rangle = \frac{I_{D}-I_{SiO_2}}{A_D}\cdot\frac{A_{SiO_2}}{I_{SiO_2}},
\label{eq:EF}
\end{equation}

\noindent
where $I_{D}$ and $I_{SiO_2}$ are the spectrally integrated PL intensity measured on each dimer nano-antenna and on the flat SiO$_2$ substrate respectively. The area of each dimer nano-antenna and the area of the laser spot are represented as $A_{D}$ and $A_{SiO_2}$ respectively. Using this definition, we calculated enhancement factors of 120, 235 and 241 for nano-antennas NA$_1$, NA$_2$ and NA$_3$ respectively when compared to a flat monolayer on the SiO$_2$ substrate. If monolayer emission from positions on bulk WS$_2$ (seen at the left edge of Fig. \ref{F2}(c)) rather than on flat SiO$_2$ is used for comparison, the enhancement factors are estimated to be 246, 486 and 533 for nano-antennas NA$_1$, NA$_2$ and NA$_3$ respectively. This is due to the lower monolayer PL on bulk WS$_2$ when compared to flat SiO$_2$, which we largely attribute to charge transfer between the WSe$_2$ and WS$_2$ crystal \cite{Lee2014}. We also compared $\left\langle EF \right\rangle$ to simulations of the fluorescence intensity enhancement in the vicinity of dimer nano-antennas with the same geometry and found agreement (see Supplementary Note 5). 

Another method of probing the photonic enhancement of the WSe$_2$ emission due to the nano-antennas is to induce a linear polarization in the excitation source and rotate this with respect to the dimer axis. As the electric field intensity surrounding the nano-antennas is higher for the X-pol mode when compared to the Y-pol mode, which can be observed in Fig. \ref{F2}(b), the PL intensity is also expected to increase for this polarization. In this experiment, the integrated intensity at each polarization angle was normalized to that from a monolayer region on flat SiO$_2$, defined as $I_{D}(\theta)/I_{SiO_2}(\theta)$, so as to highlight the effect of the nano-antenna photonic resonances. The results are shown in Fig. \ref{F2}(e) where all structures yield a degree of linear polarization of the photoluminescence, shown on the left of each plot, confirming the coupling of the emission to the nano-antenna resonances.

The last experiment performed in order to study the photonic enhancement of WSe$_2$ monolayer emission due to the WS$_2$ dimer nano-antennas was a measurement of the PL decay time, displayed in Fig. \ref{F2}(f). For this study, an avalanche photo-diode was used as a detector with an instrument response function (IRF, grey) defined by the laser pulse ($\approx$ 90 ps). A very low excitation power density of 0.02 $\mu J/cm^2$ was used in order to avoid exciton-exciton annihilation processes, which dominate in these materials at higher powers \cite{Mouri2014a}. The results yield single exponential decay lifetimes, which consist of contributions from both radiative and non-radiative recombination rates. An increase in the radiative rate due to photonic enhancement will lead to a decrease in the emission lifetime as this component is shortened. The intrinsic WSe$_2$ monolayer PL lifetime at room temperature measured from the monolayer on the WS$_2$ bulk crystal was $\tau = 1.78\pm0.08$ ns. The PL decay times measured for NA$_1$, NA$_2$ and NA$_3$ are $\tau_1 = 1.34\pm0.02$ ns, $\tau_2 = 0.96\pm0.02$ ns, $\tau_3 = 1.45\pm0.02$ ns respectively. These are lower than that measured on the bulk crystal suggesting the presence of Purcell enhancement. If we assume that the non-radiative rate contribution to the lifetime is low, we can extract Purcell factors of 1.33, 1.85 and 1.23 for NA$_1$, NA$_2$ and NA$_3$ respectively. However, previous measurements of the quantum efficiency of WSe$_2$ monolayers report values ranging from 0.06$\%$ to 5$\%$ \cite{Mohamed2017a,Kim2019,Roy2020}, therefore, the non-radiative recombination rate is much higher than the radiative. As the Purcell factor only affects the radiative component of the lifetime, the values extracted above are only lower bounds on the emission enhancement factor, which may be much higher. As suggested by the higher overlap of the PL emission with the scattering resonances of the nano-antenna, the shortest PL lifetime was measured for NA$_2$ where the highest Purcell enhancement is expected. This suggests that the enhancement factor can be modulated by tuning the dimer nano-antenna resonances closer or further from the PL emission energy of the monolayer (see Supplementary Note 5). We also varied the linear polarization of the excitation source and measured the PL decay time for the X-pol and Y-pol modes. For the majority of the measured nano-antenna sites, the X-pol mode yielded a lower lifetime than the Y-pol mode (see Supplementary Note 6) as expected from the higher simulated electric field intensities, in Fig. \ref{F2}(b), and therefore higher Purcell factors.

As further evidence of the photonic capabilities of WS$_2$ dimer nano-antennas, we performed second harmonic generation experiments using an anapole resonance present in 60 nm high nano-antennas with a radius of 205 nm and a gap of 130 nm. The anapole resonance at a wavelength of 800 nm led to confinement of the excitation laser and a 7.2 times enhanced SHG signal when compared with bulk crystal. The enhancement also proved to be polarization dependent in dimer nano-antennas, as opposed to monomers, as the electric fields are confined outside (inside) the nano-structure geometry for the X-pol (Y-pol) anapole mode. This behavior was confirmed by simulations of the confined electric energy, which also reveal that the SHG enhancement polarization orientation can be rotated with a change in excitation wavelength and its degree of linear polarization can be modulated with an increase in dimer gap (see Supplementary Note 7).\\

\noindent\textbf{AFM repositioning} The fabrication procedure outlined in Fig. \ref{F1} yields minimum dimer separation gaps of 50 nm, which result in resonances with limited electric field intensity and therefore Purcell factors. We attempted to improve upon this by using the properties of layered materials to our advantage. The relatively weak van-der-Waals forces, which facilitate mechanical exfoliation of thin WS$_{2}$ crystals also allow for a weak adhesion of the fabricated structures to the SiO$_2$/Si substrate. As a post-fabrication procedure, we employed an AFM cantilever in order to translate one nano-pillar with respect to the other and achieve gaps as small as 10 nm without damaging the nano-antenna. Smaller separation distances may be achievable; however, the accurate measurement of the gap width is a considerable challenge due to the finite size of the cantilever tip, which ranges from 1 to 10 nm. 

\begin{figure}[h!]
	\centering
  \includegraphics[width=\linewidth]{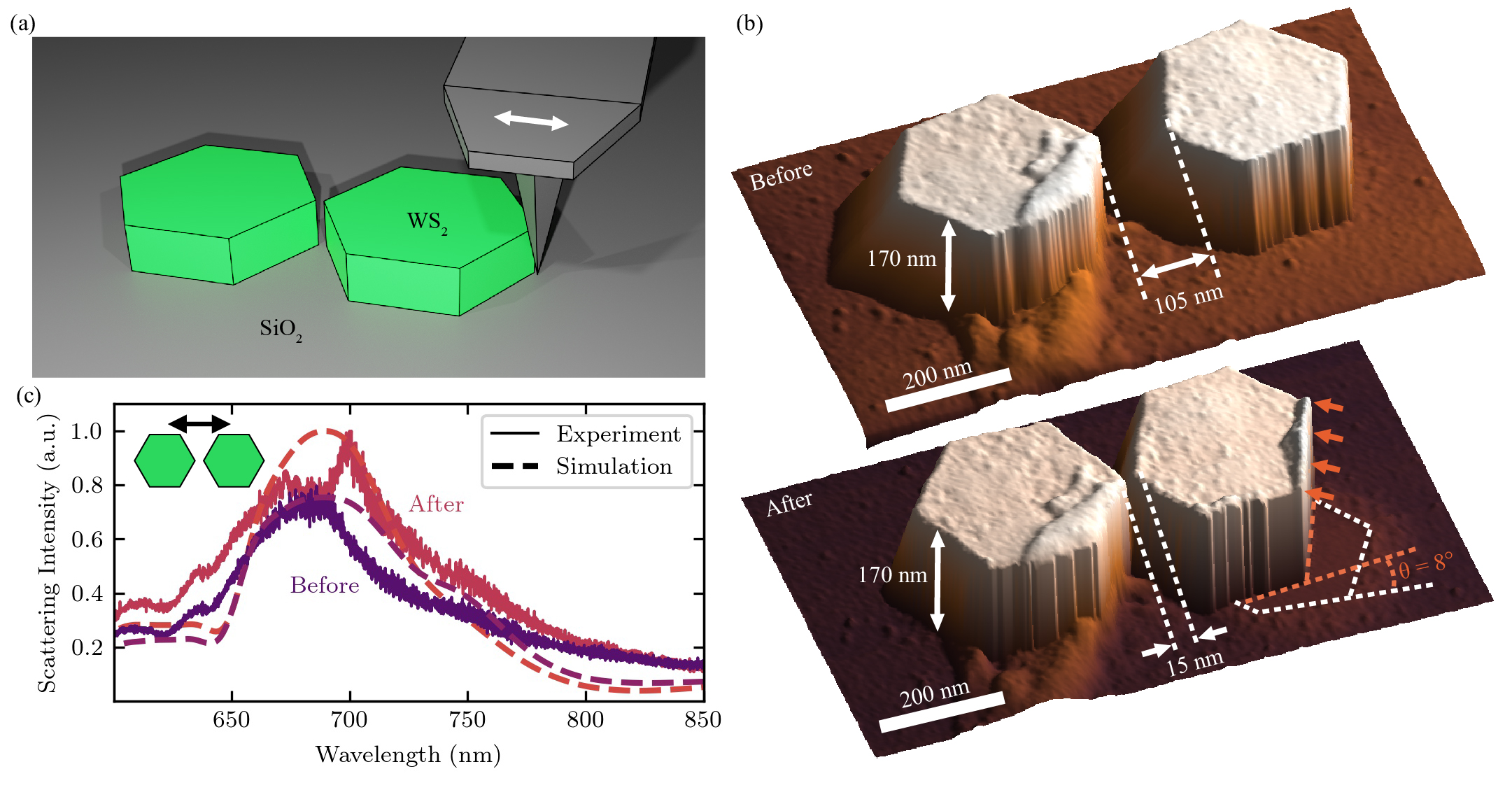}
	\caption{\textbf{AFM repositioning of dimer nano-antennas.}
	\textbf{(a)} Schematic illustration of the scanning technique used in contact mode AFM to translate and rotate single nano-pillars of the dimer nano-antenna.
	\textbf{(b)} AFM scan of a dimer structure (r = 140 nm, h = 170 nm, gap = 105 nm) before and after manipulation. The dimer gap is reduced to 15 nm and one pillar is rotated by 8$^{\circ}$ in order to position inside vertices closer to each other.
	\textbf{(c)} Dark field scattering spectra of the dimer-nano-antenna for excitation parallel to the dimer axis (as shown in inset) before and after repositioning compared to a simulated scattering cross section with gaps of 105 nm and 15 nm.}  
  \label{F3}
\end{figure}

Nano-antennas were repositioned by using AFM in contact mode. Scanning parallel to the dimer axis in a small area immediately adjacent to one nano-pillar with the scan slightly overlapping the distal edge of the nano-antenna forced the cantilever tip to displace the structure closer to the other. A schematic representation of the translation is shown in Fig. \ref{F3}(a). Very fine positioning can be achieved using this method as shown in Fig. \ref{F3}(b) where we translated one nano-pillar with respect to the other reducing the dimer gap separation from 105 nm to 15$\pm$5 nm.

An additional advantage of using this repositioning method is the ability to rotate the nano-pillars with great precision (<1$^{\circ}$). As shown in the bottom panel of Fig \ref{F3}(b), by scanning the AFM cantilever along a tangent of one of the nano-pillars and contacting with only the edge, it was rotated by over 8$^{\circ}$. Scanning closer to the midpoint of the nano-pillar translated it closer to the other. These methods of contacting the AFM tip with the nano-pillars allow for fine positioning of the inside vertices of the dimer. The repositioning procedure is iterative and entirely reversible, allowing alignment to be fine-tuned or completely changed as necessary. Further AFM scans of repositioned nano-antennas are displayed in Supplementary Note 8 confirming the reproducibility of the procedure and showing the minimum gap separation we have achieved (10$\pm$5 nm).

To confirm that the AFM repositioning modified the photonic response, we performed dark field spectroscopy before and after the translation procedure. As shown in  Fig. \ref{F3}(c), the dipole resonance seen in scattering increased after repositioning for an excitation parallel to the dimer axis, which is the most sensitive configuration to changes in gap separation. The dashed curves in the same figure represent simulations of the geometry before and after repositioning with close agreement to experiment supporting the achievement of a 15 nm gap.\\

\noindent\textbf{Electric field and Purcell enhancement of emission} Ultra-small gaps, such as those achieved with the AFM repositioning technique, are expected to provide very high electric field confinements \cite{Choi2017}. We therefore simulated the electric field intensity and Purcell enhancement induced by three dimer designs, corresponding to the different geometries shown in Fig. \ref{F1}, with a gap of 10 nm. Each design was optimized for electric field confinement at the top surface of the dimer.

\begin{figure}[h!]
	\centering
  \includegraphics[width=\linewidth]{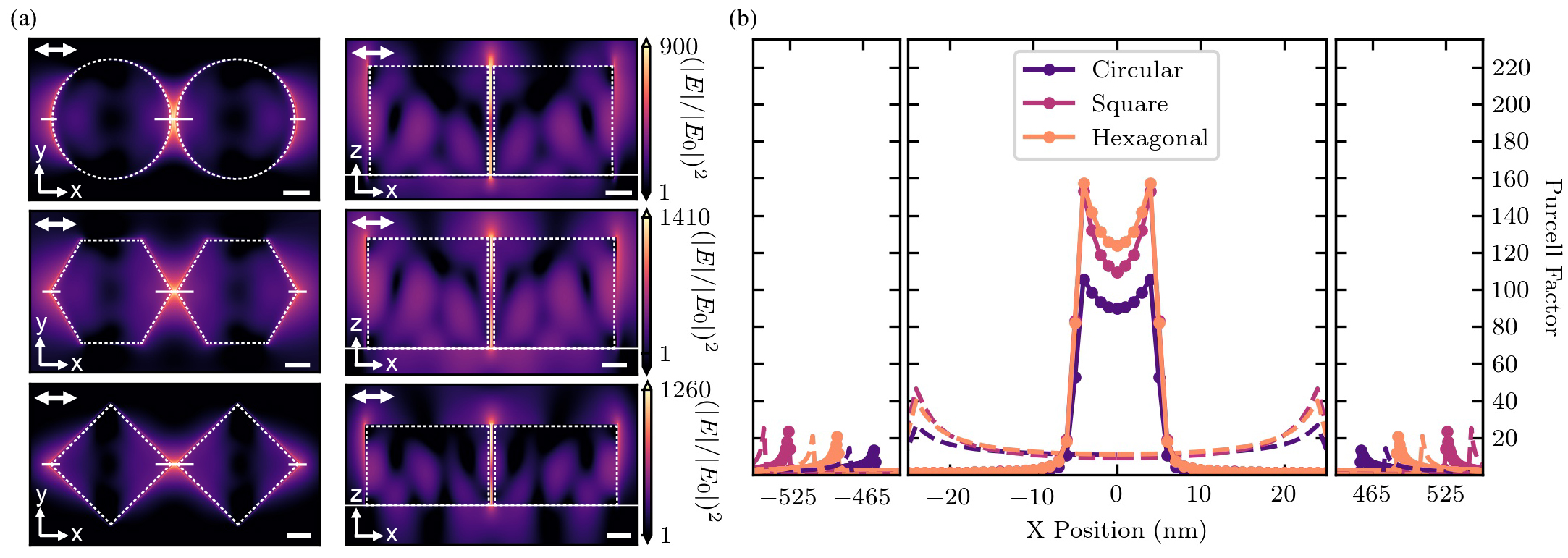}
  \caption{\textbf{Simulations of the electric field hotspots and Purcell enhancement for single photon emission.}
	\textbf{(a)} Top-view and side-view spatial distributions of the electric field intensity in and surrounding optimized designs of each geometry of WS$_{2}$ dimer nano-antennas at 10 nm separation gap. Circular design: r = 225 nm, h = 200 nm, wavelength = 751.5 nm. Hexagonal design: r = 240 nm, h = 200 nm, radius of curvature of vertices = 22 nm, wavelength = 749.5 nm. Square design: r = 260 nm, h = 150 nm, radius of curvature of vertices = 10 nm, wavelength = 697.5 nm. Hotspots of electric field confinement form at inner and outer edges of the dimer. Incident plane wave polarization (X-pol) is parallel to the double white arrow in the top left corner of the panels. Axes indicate cross-section with (x,y) indicating a surface 0.5 nm above the top of the dimer and (x,z) indicating a vertical cut through the dimer axis. Dashed white outlines represent the physical edges of the structures. Scale bars = 100 nm.
	\textbf{(b)} Purcell enhancement for a dipole placed at different positions with polarization parallel to the dimer axis, 0.5 nm above the top surface of the structures. Solid dots and curves indicate dimers with a gap of 10 nm and dashed curves indicate dimers with a gap of 50 nm. Placement positions of the dipole above the dimer are shown as solid white lines in \textbf{(a)}.}
  \label{F4}
\end{figure}

The electric field intensity spatial distribution is shown 0.5 nm above the top surface of the dimers in the left column of Fig. \ref{F4}(a), and as a cross sectional cut along the z-axis through the middle of the individual nano-pillars in the right column of Fig. \ref{F4}(a). Each distribution was calculated at the wavelength of the maximum electric field intensity (751.5 nm, 749.5 nm, 697.5 nm for the circular, hexagonal and square geometries respectively). The radii of the optimized geometries are r = 225 nm, r = 240 nm and r = 260 nm for the circular, hexagonal and square geometries respectively, while their heights are h = 200 nm, h = 200 nm and h = 150 nm. In order to closely approximate realistic structures, which can be fabricated, we measured the radius of curvature of the vertices of fabricated hexagonal and square dimer nano-antennas using atomic force microscopy. This yielded vertex radii of curvature as low as 22 nm for the hexagonal and 10 nm for the square geometry (see Supplementary Note 9). These values were then subsequently taken into account in the simulation of the hexagonal and square geometries. The electric field hotspots forming at the vertices between the two nano-pillars for an incident plane wave polarized parallel to the dimer axis exhibit intensities of more than 10$^3$ compared to vacuum as shown in Fig. \ref{F4}(a). We compare the hotspots simulated for the three geometries achievable through the fabrication process. The hexagonal and square shaped nano-antennas induce a higher electric field intensity confinement than the circular design, with the hexagonal geometry inducing the largest enhancement.

Furthermore, we evaluated the Purcell factor for a single photon source positioned onto these structures by simulating a dipole emitter at positions along the dimer axis with a polarization parallel to the same axis. It is displaced by 0.5 nm from the top surface of the structure and its emission wavelength was set to the one used for the respective geometry in Fig. \ref{F4}(a). The calculated Purcell factors at each position are shown in Fig. \ref{F4}(b). Solid white lines in the left panels of Fig. \ref{F4}(a) indicate the simulated positions of the emitter. The maxima seen at positions corresponding to the inner edges of the nano-antenna yield the largest results (>100) as expected from the electric field hotspots seen in Fig.\ref{F4}(a). Weakly confined hotspots at the outside edges of the structure also exhibit local maxima in the Purcell factor with values as high as 20. For the minimum achieved gap separation in the post-fabrication AFM repositioning, the hexagonal geometry exhibits the highest Purcell factor within its hotspot (157) with the square as a close second (153). The circularly shaped dimer leads to a maximal Purcell factor of 105. We have also simulated the Purcell factors for the minimum gap separation achievable without the use of AFM repositioning (50 nm), which are shown in Fig. \ref{F4}(b) as dashed curves. This yields much smaller Purcell enhancements (as high as 46). There are several degrees of freedom in the fabrication process that can be used to modulate both the electric field intensity and the Purcell enhancements expected for these dimer structures. One method is to vary the gap separation of the dimer, leading to an exponential decrease in the electric field intensity and Purcell factor by one order of magnitude for a gap of 100 nm when compared to a gap of 10 nm. A similar decrease in both factors can also be achieved by rotation of one nano-pillar with respect to the other (see Supplementary Note 10). Such modulations can be experimentally achieved through the use of nano-fabrication techniques or AFM repositioning.\\

\begin{figure}[h!]
	\centering
  \includegraphics[width=\linewidth]{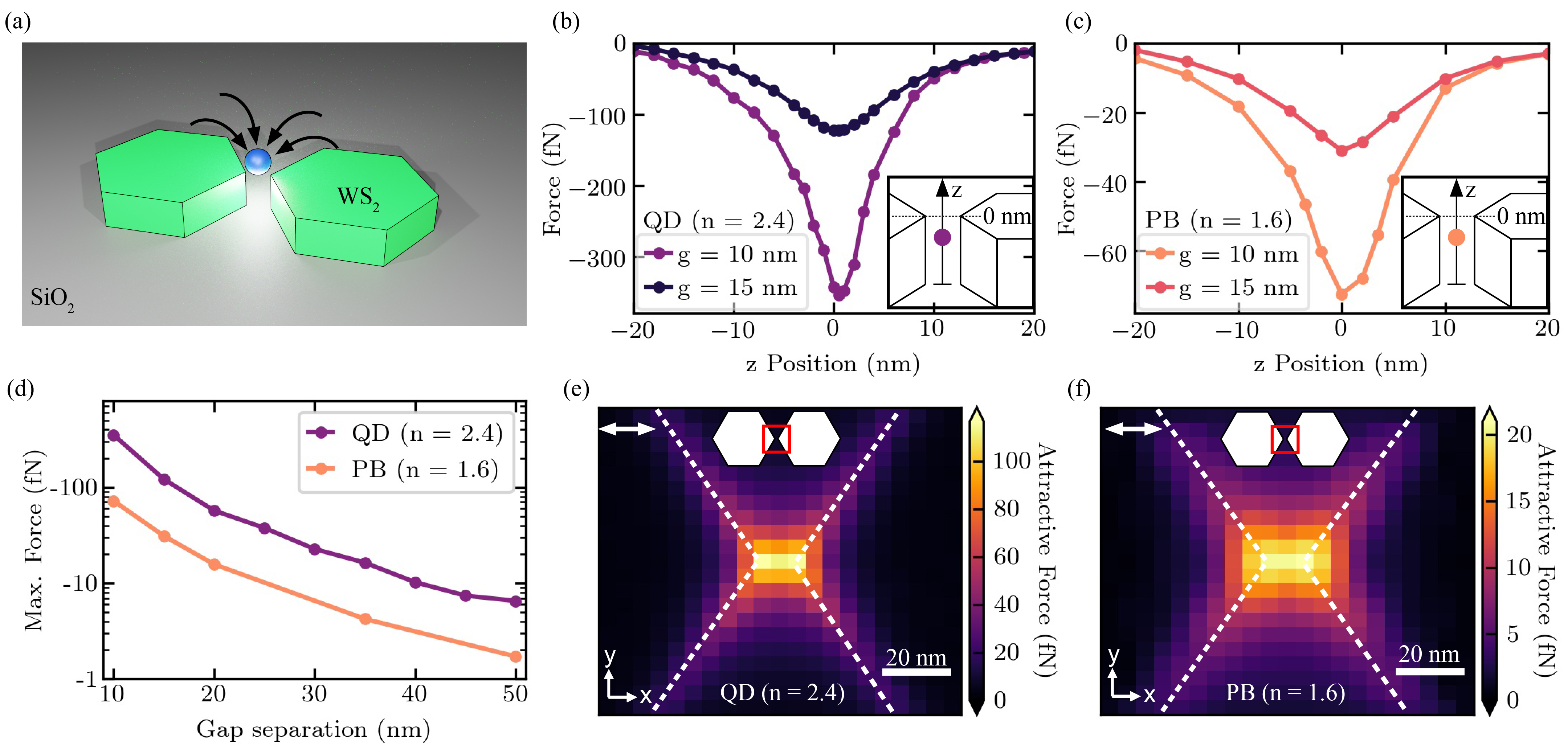}
	\caption{\textbf{Optical trapping simulations of nano-particles using WS$_{2}$ hexagonal nano-antennas.}
	\textbf{(a)} Schematic illustration of optical trapping of a nano-particle in water through the use of WS$_{2}$ hexagonal dimer nano-antenna hotspots.
	\textbf{(b), (c)} Comparison of the optical force for a spherical nano-particle (r = 5 nm) placed in the middle of the dimer nano-antenna (r = 240 nm, h = 200 nm) gap, along the z-axis for gaps of 10 nm and 15 nm. The trapping force is calculated at the wavelength of its maximum value for a QD \textbf{(b)} (763 nm) and a PB \textbf{(c)} (755 nm). Insets illustrate the positioning of the respective nano-particle with the zero position at the top surface of the nano-antenna.
	\textbf{(d)} Maximum electric force applied to a QD (purple) and a PB (yellow) at the top surface of a dimer nano-antenna (r = 240 nm, h = 200 nm) with a varying gap separation. 
	\textbf{(e), (f)} Attractive force distribution for a nano-particle in a plane 5 nm from the top surface of a dimer nano-antenna (r = 240 nm, h = 200 nm, g = 10 nm) centered at the gap. The trapping force is calculated at the wavelength of its maximum value for a QD \textbf{(e)} (763 nm) and a PB \textbf{(f)} (755 nm). The double white arrow in the upper left portion of the figure identifies the pump polarization along the dimer axis (X-pol). Dashed white lines identify the physical edges of the structure. Inset schematics highlight the position of the plane over which the optical force is simulated. Maximum force values coincide with the electric field hotspots. Input pump power density used for all simulations is 10 $mW/\mu m^2$.}
  \label{F5}
\end{figure}

\noindent\textbf{Dimer nano-antenna optical trapping} In order to study the potential of ultra-small gap WS$_{2}$ dimer nano-antennas for optical trapping, we perform numerical simulations based on the Finite Element Method (FEM) to determine the Maxwell Stress Tensor (MST) and calculate the force exerted on small dielectric particles. The simulation geometry consists of the earlier optimized hexagonal nano-antenna with the addition of a nano-sphere (r = 5 nm) with a refractive index corresponding to either an approximated colloidal quantum dot (n = 2.4) \cite{Xu2019} or a polystyrene bead (n = 1.6), which closely mimics the refractive index and size of a large protein \cite{Young2017}. Since optical trapping experiments often require a suspension of the nano-particles in a solution, we set the background refractive index to that of water. For all simulations, we use an experimentally feasible pump power density of 10 $mW/\mu m^2$. A schematic representation of the simulation is shown in Fig. \ref{F5}(a).

We explore the optical force applied to the QD and the PB, shown in Fig. \ref{F5}(b) and (c) respectively, as they are translated along the z-axis in the middle of the dimer gap. We simulate the forces for two gap separations of the dimer hexagonal nano-antenna (10 nm and 15 nm) and observe an attractive (negative) force, which is maximized at the top surface of the nano-antenna where the electric field hotspots are formed. We observe a maximum optical force of 353 fN on the QD and 73 fN on the PB for the dimer nano-antennas with a 10 nm separation. For the larger dimer gap (15 nm), this reduces to 123 fN on the QD and 31 fN on the PB. We also study the dependence of this maximum optical force in the electric field hotspots on the dimer gap, which is varied from 10 nm to 50 nm. As shown in Fig. \ref{F5}(d), the maximum attractive force imparted on the nano-particles decreases exponentially by one order of magnitude as expected from the decay of the electric field intensity with an increased separation (see Supplementary Note 10). 

We subsequently explore the lateral spatial dependence of the optical force by simulating the QD and PB in a plane, which is 5 nm above the top surface of the dimer nano-antenna with a 10 nm gap, as shown in Fig. \ref{F5}(e) and (f). This is an emulation of the position of nano-particles placed onto the top surface of the dimer. The optical force, as expected, is maximized at the position of the electric field hotspots yet still maintains values of more than 100 fN for the QD and 20 fN for the PB.\\

\large
\noindent\textbf{Discussion}\\
\normalsize

We have fabricated and characterized both monomer and dimer WS$_{2}$ nano-antennas with the aim to highlight the advantages of their use in photonic applications such as tunable photoluminescence enhancement, polarization dependent SHG enhancement, SPE enhancement and optical trapping thereby broadening the versatility of TMD nano-antennas. Dark field studies on monomer and dimer nano-antennas reveal a straightforward approach for tuning resonances in the structures in order to fit with different applications by varying the radius, height or geometry. We couple monolayer WSe$_2$ emission to the dimer resonances in the same, TMD, material system and achieve PL enhancement factors of more than 240 and a Purcell enhancement factor lower bound of nearly 2. Through the use of a dimer anapole mode we show polarization dependent SHG enhancement, not present in bulk TMD flakes or monomer nano-antennas, that can be rotated by a change in excitation polarization which may provide advantageous for the realization of optical logic gates \cite{Bovino2009}. Post-fabrication repositioning of dimer nano-antennas utilizing AFM led us to attain ultra-small gaps of 10$\pm$5 nm, on the order of the feature resolution limit set by FIB milling \cite{Novotny2011}, which we achieve through a more precise, less damaging technique. Although a previous report has demonstrated AFM positioning of plasmonic bowtie nano-antennas \cite{Merlein2008}, this technique has only become possible for dielectric nano-antennas made from layered materials due to their intrinsic van-der-Waals attractive forces. This allows us to achieve the smallest separation recorded for a dielectric resonator defined through the use of electron beam lithography followed by reactive ion etching\cite{Regmi2016,Xu2018}. Another advantage of this method is the possibility to change the relative orientation of individual nano-pillars, thereby, aligning their potentially atomically sharp vertices in closer or further proximity.

The numerical simulation studies of dimer nano-antennas exhibiting an ultra-small gap yield highly confined electric field intensity hotspots (10$^3$ enhancement compared to vacuum). We predict the practical utility of WS$_{2}$ nano-antennas for radiative rate enhancement of single photon emission with Purcell factors of up to 157 for a hexagonal and 153 for a square geometry. The expected quantum emission enhancement in the dimer nano-antennas is higher than the largest currently achieved in photonic crystal cavities \cite{Liu2018b} due to the previously inaccessibly small proximity attained with the AFM repositioning method. We also explore two different routes to modulate the photonic enhancement through variation of the dimer gap as well as the relative rotation of the individual nano-pillars, which are both controllable either through the fabrication process or, to a greater precision, the post-fabrication repositioning. These methods of controlling the emission properties of single photon sources may prove useful for WSe$_{2}$ SPEs, which form at high strain gradients in monolayers transferred onto dimer nano-antennas collocated with the electric field hotspots \cite{Sortino2020}. Previous reports have shown quantum efficiency enhancement for WSe$_{2}$ SPEs \cite{Sortino2021} as well as rotation of the emitter dipole moment due to a change in the strain gradient \cite{Kern2016} both of which can be controlled by a change in separation and rotation of individual nano-pillars in the dimer nano-antenna through AFM repositioning. 

We further explored a route to the precise positioning of SPEs by simulating optical trapping forces due to the highly confined electric field in the gap of the dimer nano-antennas. We calculated a maximum attractive force of 353 fN for a colloidal quantum dot and 73 fN for a protein-like, polystyrene bead both with a radius of 5 nm under a pump power density of 10 $mW/\mu m^2$. When compared to previous examples of dielectric nano-antennas for optical trapping, the WS$_{2}$ dimers yield higher attractive forces by a factor of >83 for QDs \cite{Xu2019} and >40 for PBs \cite{Xu2018} with the same size and under the same pump power conditions. Therefore, WS$_{2}$ dimer nano-antennas with ultra-small gaps show great potential for applications of stable trapping of very small nano-particles with a moderate optical power. This once again highlights the advantage of the AFM repositioning technique to reduce the dimer gap below the limits available to standard nano-fabrication. The large Purcell enhancements and optical trapping forces predicted for WS$_{2}$ dimer nano-antennas with ultra-small gaps highlight the unique potential for TMD nano-resonator research and applications. This is possible due to the refractive index and van-der-Waals forces, which allow the formation of highly confined resonances and hotspots while simultaneously opening the possibility of AFM repositioning. The field of nano-photonics includes a diverse library of materials and here we assert and expand the possibilities provided by adding thin crystals of TMDs to the list.\\

\large
\noindent\textbf{Methods}\\
\normalsize

\noindent\textbf{Sample fabrication}

\noindent\textit{WS$_2$ exfoliation}: WS$_{2}$ flakes were mechanically exfoliated from bulk crystal (HQ-graphene) onto a nominally 290 nm SiO$_{2}$ on silicon substrate. Large flakes with recognizable crystal axes via straight edged sides at 120$^{\circ}$ to each other were identified and their positions within the sample were recorded for further patterning.

\noindent\textit{Electron beam lithography}: Samples were spin coated with ARP-9 resist (AllResist GmbH) at 3500 rpm for 60 s and baked at 180$^\circ$ for 5 min yielding a film of 200 nm thickness. Electron beam lithography was performed in a Raith GmbH Voyager system operating at 50 kV using a beam current of 560 pA.

\noindent\textit{Reactive ion etching}: Anisotropic etching to imprint the resist pattern into the WS$_{2}$ flakes physically was carried out using a mixture of CHF$_{3}$ (14.5 sccm) and SF$_{6}$ (12.5 sccm) at a DC bias of 180 V and a pressure of 0.039 mbar for 40 seconds. Isotropic etching was achieved by using a more chemical recipe with solely SF$_{6}$ (20 sccm) at a DC bias of 50 V and a pressure of 0.13 mbar for 40 seconds. Removal of the remaining resist after etching was accomplished by a bath in warm 1165 resist remover (1 hour) followed by Acetone (5 min) and IPA (5 min). If resist is still found on the sample, final cleaning is done in a bath of Acetone (1 hour) and IPA (5 min) followed by 1 hour in a UV ozone treatment. In some cases, the structures were slightly overetched leading to nano-antennas with a small pedestal of SiO$_2$ (<20 nm). This, however, did not lead to any noticeable changes in the photonic resonances nor in the ability to reposition the structures with AFM.

\noindent\textit{WSe$_2$ transfer}: WSe$_{2}$ monolayers were mechanically exfoliated from a bulk crystal (HQ-graphene) onto a (PDMS) stamp, which had previously been attached to a glass slide. Large monolayers were identified using PL imaging. The glass slide is rotated upside down and attached to a holder arm by means of a vacuum. The target substrate, consisting of WS$_2$ nano-antennas on a SiO$_2$ surface, was also held to a stage using the same vacuum. The WSe$_2$ monolayer was slowly brought into contact with the target substrate through the use of a piezo-scanner stage. After the entire monolayer has contacted the surface, the glass slide with PDMS was slowly moved away from the target substrate. The low speed of the peeling process makes use of the visco-elastic properties of the PDMS polymer and leaves the monolayer of WSe$_2$ onto the substrate.\\

\noindent\textbf{Dark field spectroscopy}
Optical spectroscopy in a dark field configuration was achieved using a Nikon LV150N microscope with a fiber-coupled output. Incident illumination from a tungsten halogen lamp in the microscope was guided to a circular beam block with a diameter smaller than the beam diameter. The light was then reflected by a 50:50 beam-splitter towards a 50x Nikon (0.8 NA) dark-field objective which only illuminates the sample at large angles to the normal. Reflected light from the sample is guided back through the same objective towards a fiber coupler. Due to the small diameter of the multi-mode fiber core used, only light reflected back at small angles to the normal is collected. The fiber from the microscope was subsequently coupled to a Princeton Instruments spectrometer and charge coupled device (CCD).\\

\noindent\textbf{Micro-Photoluminescence spectroscopy}
In order to record the photoluminescence emitted from monolayer WSe$_2$ at different regions of our sample, we used a home-built setup, which includes a pulsed diode laser at 638 nm. The sample was mounted into an Oxford Instruments flow cryostat and the chamber was pumped to vacuum. The collimated excitation laser was passed through a 700 nm short-pass filter, a Glan-Thompson linear polarizer and a half wave plate before being deflected by a 50:50 beam-splitter and passing through a 100x (0.7 NA) Mitutoyo objective which focused the beam onto the sample. The emitted light is collected by the same objective and passes through the beam-splitter to be guided through a 700 nm long-pass filter and is focused onto the slit of a Princeton Instruments spectrometer (0.75 meter) and CCD (data shown in Figs. \ref{F2}(d) and (e) and Supplementary Note 5 and 6). The PL decay time studies utilized the pulsed laser excitation and the emitted light was spectrally filtered (10 nm) using the exit slit of the spectrometer before it was fiber-coupled to an ID Quantique avalanche photo-diode (id100) (data shown in Fig. \ref{F2}(f) and Supplementary Note 6).\\
 
\noindent\textbf{Atomic force microscope repositioning and imaging}
All repositioning was carried out using a JPK Nanowizard 3 Ultra AFM using Bruker SNL probes (cantilever C, nominal stiffness 0.24 N/m). First, nano-antennas were imaged in QI mode with a setpoint of 1 nN, a Z length of 400 nm and a pixel time of 15 ms. To reposition the nano-antennas, the AFM was switched to contact mode and 200 nm scans were performed with a setpoint of 1 nN and a scan rate of 2 Hz on the substrate immediately adjacent to the nano-antenna. The fast scan axis was oriented along the desired direction of movement and the scan area was progressively moved so that the scan overlapped with the nano-antenna to translate it in the desired direction in increments of 5-50 nm at a time. Periodically, the nano-antenna was re-imaged in QI mode to check the relative position and orientation of the pillars. Final characterization after repositioning was performed in QI mode.\\

\noindent\textbf{FDTD simulations}
The finite-difference time-domain simulations were carried out using Lumerical Inc. software.

\noindent\textit{Scattering simulations}: Calculations of the scattering cross section shown in Fig. \ref{F3} and Supplementary Notes 1 and 3 were carried out by defining the geometry of the WS$_{2}$ nano-antennas onto a SiO$_{2}$ substrate utilizing the refractive index of WS$_{2}$ from reference \cite{Verre2019}. Illumination with a plane wave was sent normal to the surface using a TFSF source from the air side. The illumination was unpolarized, however, for Fig. \ref{F3} the polarization was set parallel to the dimer axis. The scattered intensity was subsequently collected from a monitor set above the illumination plane (in the far-field) so that the dark field spectroscopy experiments could be closely emulated.

\noindent\textit{Electric field intensity simulations}: Calculations of the near-field electric field intensity normalized to vacuum, shown in Figs. \ref{F2}(b) and \ref{F4}(a) were simulated using the same geometry and illumination scheme as for the scattering simulations (polarization is shown as a white double arrow in top left corner of each panel) with a monitor recording the electric field 0.5 nm above the top surface of the nano-antennas or as a vertical cross-section of the structure passing through the dimer axis as shown in the right-most panels of Fig. \ref{F4}(a). For Supplementary Note 7, the monitor was designed as a cross-sectional profile through the midpoint of the nano-antenna height. For Figs. \ref{F2}(b) and \ref{F4}(a) this monitor was defined to encompass the entire cross section of the nano-antennas. For the simulations performed in Supplementary Note 5 and 10, only a single point 1 nm from the inside edge of the dimer nano-antenna within the hotspot was recorded.

\noindent\textit{Purcell factor simulations}: Simulations of the Purcell factor were carried out using the same geometry as for the electric field intensity simulations. The illumination was achieved through a dipole source placed at different positions, 0.5 nm above the top surface of the nano-antenna with a polarization parallel to the dimer axis for Fig. \ref{F4}(b). For the simulations carried out in Supplementary Note 6, the polarization was set parallel and perpendicular to the dimer axis in order to compare the two. For the simulations displayed in Supplementary Note 10 however, the position was set to 0.5 nm above the top surface of the nano-antenna and 1 nm away from the inside edge within the electric field hotspot.\\

\noindent\textbf{Optical trapping force simulations}: We have used the 3D finite element method (COMSOL Multiphysics) to calculate the optical forces in the hexagonal dimer nano-antenna. The structure is illuminated with a plane wave propagating in a normal direction to the top surface of the structure with a polarization along the dimer axis. The background refractive index was set to that of water to enable access of the bead to the trapping site. We have calculated the optical forces at the resonance of the nano-antenna, corresponding to the maximum energy enhancement. The value of the optical force is obtained by integrating the Maxwell Stress Tensor on the surface of the target nano-particle \cite{Conteduca2017}. A fine mesh (resolution of $\leq$ 3 nm) has been employed in the dimer gap to calculate the electromagnetic distribution with a high accuracy, minimizing the error on the evaluation of the optical force. The computational domain has been set to a sufficiently large value (>2 $\mu m$) and surrounded by perfectly matched layers in order to avoid undesired reflection and scattering from the boundary.\\

\large
\noindent\textbf{Acknowledgments}\\
\normalsize

P. G. Z., L. S., T.S.M., M.S., A. G. and A. I. T. thank the financial support of the European Graphene Flagship Project under grant agreements 881603 and EPSRC grants EP/S030751/1, EP/V006975/1 and EP/P026850/1. L. S. and A. I. T. thank the European Union's Horizon 2020 research and innovation programme under ITN Spin-NANO Marie Sklodowska-Curie grant agreement no. 676108. P. G. Z. and A. I. T. thank the European Union's Horizon 2020 research and innovation programme under ITN 4PHOTON Marie Sklodowska-Curie grant agreement no. 721394. T.F.K. acknowledges the support of the Engineering and Physical Sciences Research Council (grant number EP/P030017/1). Y.W. acknowledges a Research Fellowship (TOAST) awarded by the Royal Academy of Engineering. We would also like to thank Cynthia Vidal for her contribution to the photoluminescence measurements of monolayer WSe$_2$ on WS$_2$ dimer nano-antennas.\\

\large
\noindent\textbf{Author contributions}\\
\normalsize

P.G.Z and L.S. exfoliated WS$_{2}$ layers onto SiO$_{2}$ substrates. Y.W. fabricated nano-antenna structures using EBL and RIE. P.G.Z and Y.W. performed AFM and SEM characterization of fabricated nano-antennas. P.G.Z. carried out dark field spectroscopy measurements, micro-PL measurements, scattering cross section simulations and electric/magnetic field intensity profile simulations for identification of WS$_{2}$ nano-antenna resonances. P.G.Z., T.S.M. and A.G. performed second harmonic generation experiments on WS$_{2}$ nano-antennas. N.M. repositioned dimer nano-antennas using AFM. P.G.Z. and M.S., simulated electric field intensity profiles and Purcell factors for optimized dimer nano-antennas as well as for studies on dimer gap separation and rotation. D.C. performed optical trapping simulations and analyzed the results with Y.W., T.F.K. P.G.Z., L.S., T.S.M. and A.I.T. analyzed various optical spectroscopy data. J.H., T.F.K., Y.W. and A.I.T. managed various aspects of the project. P.G.Z. and A.I.T. wrote the manuscript with contributions from all co-authors. P.G.Z., L.S., Y.W., N.M., D.C., J.H., A.I.T. and T.F.K. conceived the experiments and simulations. A.I.T. oversaw the entire project.\\

\bibliographystyle{unsrt}
\bibliography{./library}

\end{multicols}

\newpage
\newpage
\newpage
\newpage

\end{document}


\pagenumbering{arabic}

\begin{center}
{\Large{\bf {\Large{\bf Supplementary Information for: Transition metal dichalcogenide dimer nano-antennas with ultra-small gaps}}}}
\vskip1.0\baselineskip{Panaiot G. Zotev$^{a*}$, Yue Wang$^{b*}$, Luca Sortino$^{a,c}$ Toby Severs Millard$^a$, Nic Mullin$^a$, Donato Conteduca$^b$, Mostafa Shagar$^a$, Armando Genco$^a$, Jamie K. Hobbs $^a$, Thomas F. Krauss$^b$, Alexander I. Tartakovskii$^{a***}$}
\vskip0.5\baselineskip\footnotesize{\em$^a$Department of Physics and Astronomy, University of Sheffield, Sheffield S3 7RH, UK\\$^b$Department of Physics, University of York, York, YO10 5DD, UK\\$^c$Chair in Hybrid Nanosystems, Nanoinstitute Munich, Faculty of Physics, Ludwig-Maximilians-Universit\"{a}t, M\"{u}nchen, 80539, Munich, Germany\\}
$^{*}$p.zotev@sheffield.ac.uk \quad $^{**}$yue.wang@york.ac.uk \quad $^{***}$a.tartakovskii@sheffield.ac.uk

\end{center}
\vskip1.0\baselineskip


\setcounter{figure}{0}
\renewcommand{\thefigure}{S\arabic{figure}}

\begin{center}
\Large
\textbf{Supplementary Note 1: Photonic resonances of monomer nano-antennas}
\end{center}

\begin{figure}[H]
	\centering
  \includegraphics[width=\linewidth]{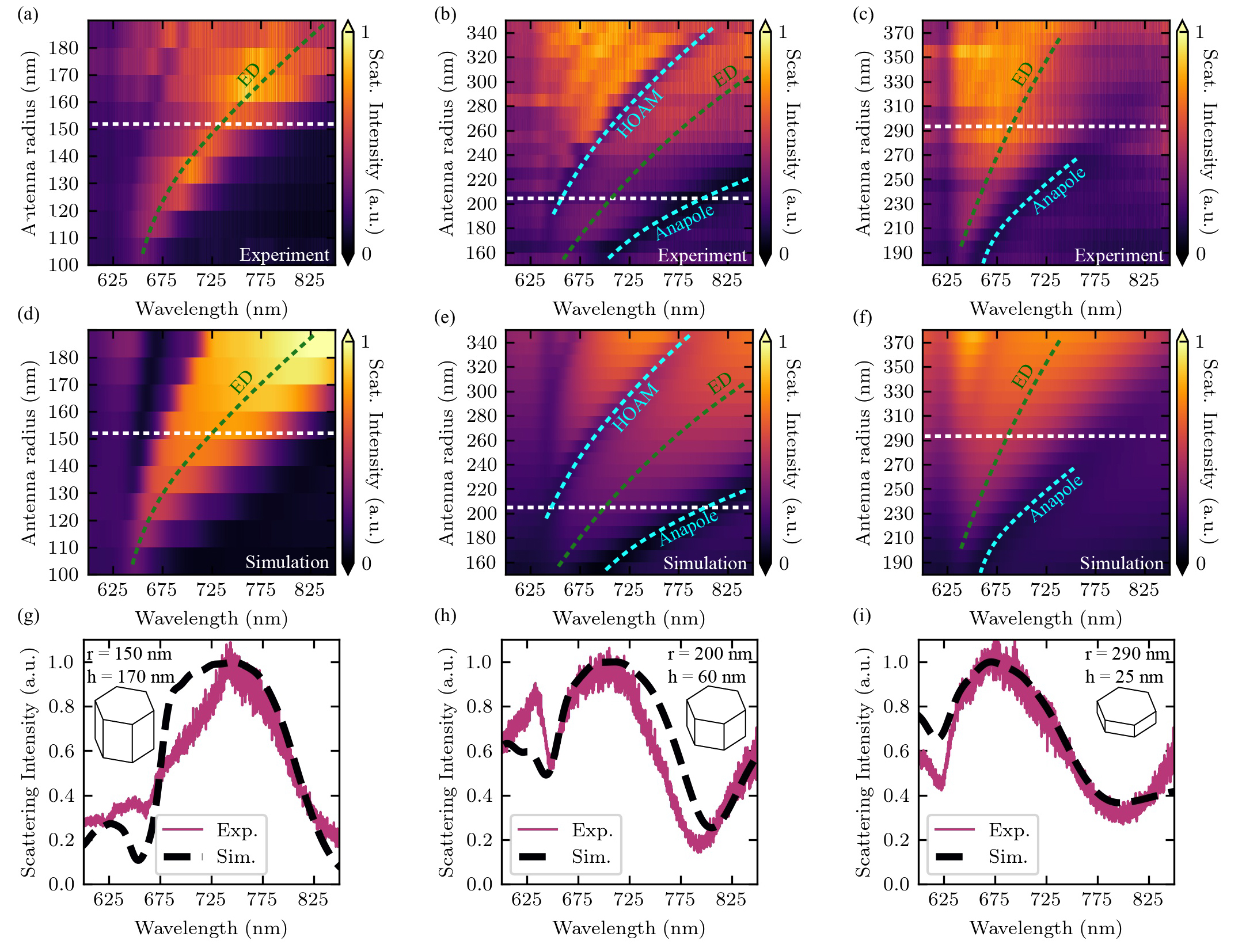}
  \caption{\textbf{Comparison of dark field spectroscopy to scattering cross section simulations for monomer nano-antennas.}
	\textbf{(a)-(c)} Dark field spectroscopy results for nano-antennas with heights of 170 nm, 60 nm and 25 nm respectively. Identified resonances include a broadened electric dipole (ED) resonance as well as anapole and higher order anapole modes (HOAM). 
	\textbf{(d)-(f)} Scattering cross section simulations for nano-antennas with heights of 170 nm, 60 nm and 25 nm respectively. Simulations confirm the identification of ED, anapole and HOAM resonances.
	\textbf{(g)-(i)} A comparison between dark field spectroscopy and scattering cross section simulations for individual radii (150 nm, 200 nm, 290 nm) as defined by a horizontal dashed white line in upper panels for the three heights (170 nm, 60 nm, 25 nm).}
  \label{MonoScat}
\end{figure}

As displayed in Fig. \ref{MonoScat}(a)-(c), we measured dark field spectra for hexagonal monomer nano-antennas of three different heights (170 nm, 60 nm and 25 nm) over a range of radii (100 - 370 nm) and subsequently simulated the fabricated geometries which, as shown in Fig. \ref{MonoScat}(d)-(f), exhibit close agreement to the experimental data. Examples of individual experimentally measured and simulated spectra for hexagonal nano-antennas of different heights are shown in Fig. \ref{MonoScat}(g)-(i). We identified an electric dipole resonance with small contributions from higher order modes (labeled ED) \cite{Hinamoto2020}. More complex resonances were also present in the fabricated monomer nano-antennas such as the anapole and higher order anapole mode (HOAM,cyan), seen as minima in dark field spectra shown in the middle and right column of Fig. \ref{MonoScat} \cite{Miroshnichenko2015}.

It is clear from the data and simulation results for all heights in Fig. \ref{MonoScat}, as expected from Mie theory, that an increase in resonator radius leads to the scattering of longer wavelengths of light, thereby red-shifting the modes. The height of the nano-antennas also plays an important role in the resonance wavelength as shown in the comparisons of experiment to simulation (Fig. \ref{MonoScat}(g)-(i)) for a single radius at each height corresponding to the dashed white lines in the upper panels. Here it can be observed that a decrease in height (from 170 nm to 25 nm) blue-shifts the resonances in the nano-antenna despite the increase in radius (150 nm to 290 nm) when moving from Fig. \ref{MonoScat}(g) to (i).

\begin{center}
\Large
\textbf{Supplementary Note 2: Scattering comparison of nano-antenna geometry}
\end{center}

The definition of the radius leaves the possibility of a change in volume when there is a change in geometry even if the radius remains the same. Since the modes inside our structures are Mie resonances, they are heavily dependent on the geometry and the volume of the high refractive index nano-antenna \cite{Mie1908}. As the volume increases so does the wavelength of the resonance because its fundamental mode must now fit inside a larger geometry. 

\begin{figure}[H]
	\centering
  \includegraphics[width=\linewidth]{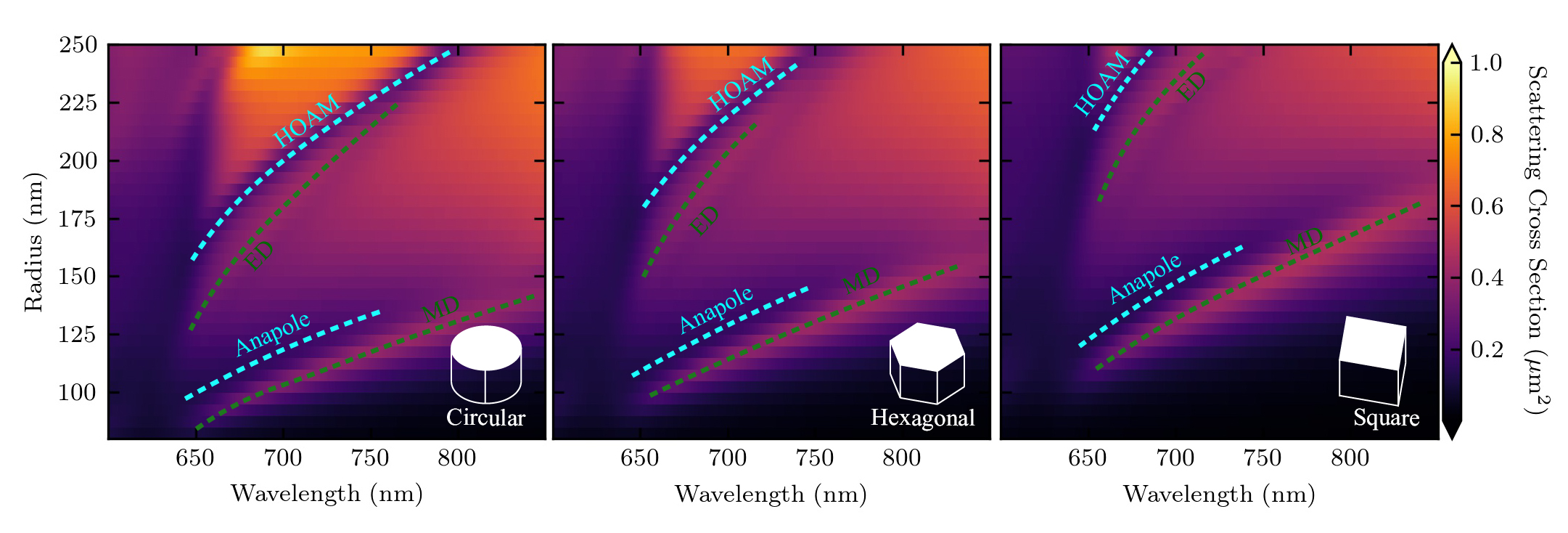}
  \caption{\textbf{Scattering cross section comparison of circular, hexagonal and square geometries.} The height of all structures was kept at 150 nm with a varying radius (80-250 nm) in order to show the blue-shift of the resonances with decreasing volume of the antenna for the hexagonal and square geometries. Identified resonances include a magnetic dipole resonance (MD), an electric dipole resonance (ED), an anapole mode and a higher order anapole mode (HOAM).}
  \label{GeoComp}
\end{figure}

In figure \ref{GeoComp} we have compared the simulated scattering cross sections of the three different WS$_{2}$ monomer nano-antenna geometries we were able to fabricate. These calculations are done in vacuum in order to allow us to identify modes, which will broaden with the addition of a substrate. We set the same height (150 nm) and range of radii (80-250 nm) for the different geometries. We are able to identify a magnetic dipole resonance, which we denote as MD, an electric dipole resonance denoted as ED and the more complex anapole and higher order anapole mode. We also notice that the resonances blue-shift when moving from circular to hexagonal to square nano-antennas with the same radius as expected from the geometrical argument. 

\begin{center}
\Large
\textbf{Supplementary Note 3: Dark field spectra of dimer nano-antennas for a range of radii}
\end{center}

\begin{figure}[H]
	\centering
  \includegraphics[width=0.983\linewidth]{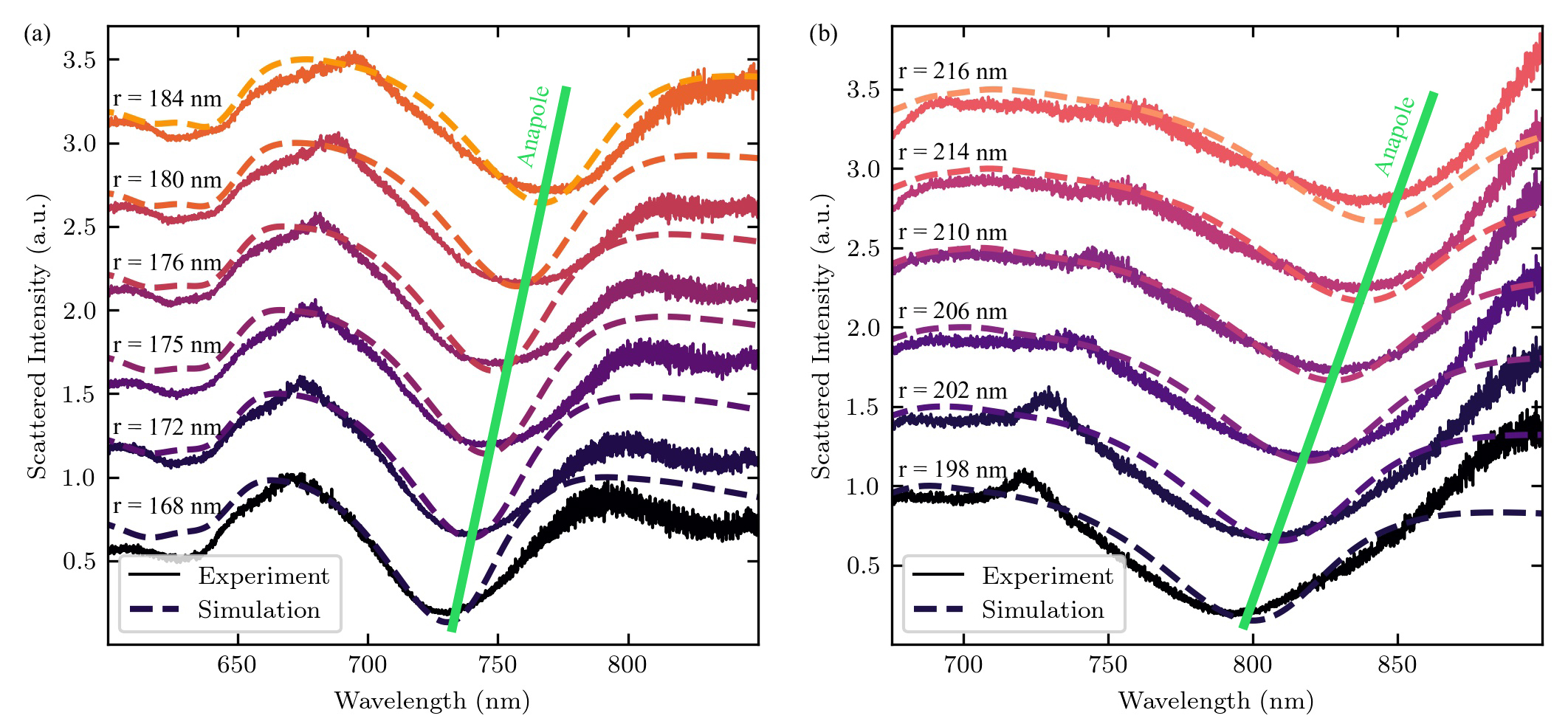}
  \caption{\textbf{Dark field scattering spectra for unpolarized excitation compared to scattering cross section simulations.} 
	\textbf{(a)} Vertically offset dark field spectra compared to simulations for a range of radii (168 nm to 184 nm) with a height of 60 nm. 
	\textbf{(b)} Same as \textbf{(a)} for a range of radii (198 nm to 216 nm) with the same height. The solid green line highlights the position of the anapole mode in these structures.}
  \label{DimScat}
\end{figure}

We recorded dark field scattering spectra, under unpolarized excitation, for WS$_2$ dimer nano-antennas shown here in Fig. \ref{DimScat}. The heights of the nano-antennas are all 60 nm with the radii ranging from 168 nm to 216 nm. The anapole mode, seen as the major feature of these spectra, red-shifts from 730 nm to 843 nm. This study provides evidence that the resonances of the dimer nano-antennas are just as tunable as for the monomers with a change in radius.

\begin{center}
\Large
\textbf{Supplementary Note 4: Hybridization of monomer nano-antenna resonances in dimer nano-antennas}
\end{center}

When two monomer nano-antennas are separated by a distance smaller or on the order of the wavelength of the individual resonances of structures, these modes begin to hybridize and therefore split energetically. This allows the formation of two new cross-polarized modes (X-pol and Y-pol) which are photonic resonances of the new dimer nano-antenna.

\begin{figure}[H]
	\centering
  \includegraphics[width=0.486\linewidth]{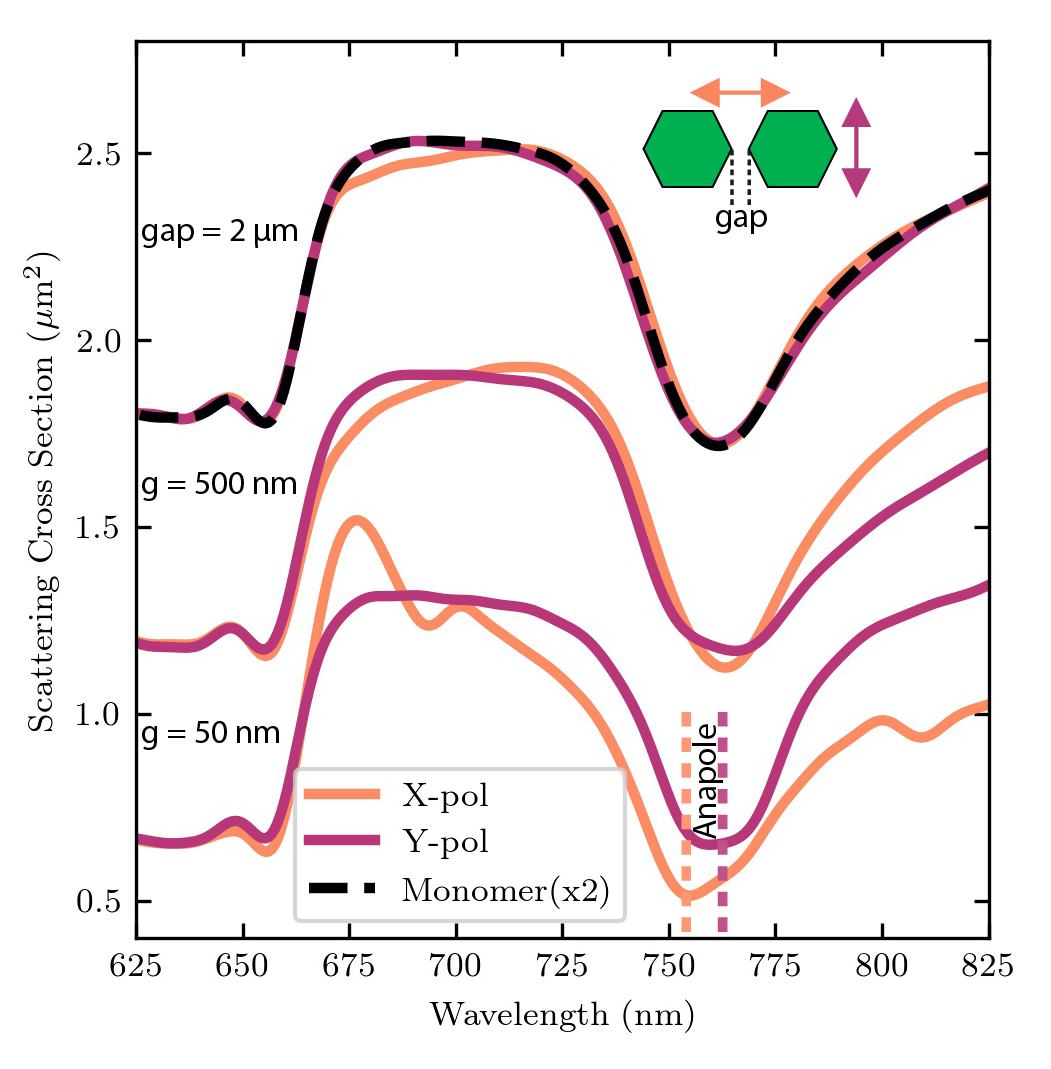}
  \caption{\textbf{Hybridization of individual resonances in dimer nano-antennas.} Simulated scattering cross sections for dimers with three different gap distances and an excitation polarization along the dimer axis (yellow) and perpendicular to it (red). At large gap distances such as 2 $\mu$m one can compare the resonances for both incident polarizations to two isolated monomer nano-antennas. At small gap distances, the modes split along the different polarization directions yielding two cross-polarized anapole and dipole resonances. Radius = 250 nm. Height = 150 nm. Spectra are vertically offset for better visualization of the results.}
  \label{Hybrid}
\end{figure}

As shown in figure \ref{Hybrid}, we can compare the simulated scattering cross section of two nano-pillars (r = 250 nm, h = 150 nm) on a SiO$_{2}$ substrate separated by a 2 $\mu$m gap to the scattering cross section of two completely isolated monomer nano-antennas. Here we simulate an excitation plane wave polarized along the dimer axis (X-pol) as shown by the yellow double arrow in the top right of the figure as well as perpendicular to this (Y-pol) as shown by the red double arrow in the inset. For a separation of 2 $\mu$m, we do not observe a splitting into perpendicularly polarized modes but there is good agreement with the scattering cross section of two individual monomer nano-antennas. However, as we reduce the separation distance of the dimer to 500 nm and subsequently to 50 nm, a splitting appears at the anapole mode minimum (750-770 nm) and at the dipole resonance (670-700 nm) for the two cross polarized modes. This clearly shows the hybridization of the individual resonances of the monomer nano-antennas placed in close proximity with relation to the wavelength of the resonances.

\begin{center}
\Large
\textbf{Supplementary Note 5: Photoluminescence enhancement of WSe$_2$ monolayer due to coupling to WS$_2$ dimer nano-antenna photonic modes}
\end{center}

In order to understand the large photoluminescence (PL) enhancement measured experimentally, we must first describe the contributing factors, which lead a photonic resonance to enhance the fluorescence of emitters such as excitons found in a two-dimensional WSe$_2$ layer. The fluorescence rate enhancement ($F/F_0$) can be written as follows:

\begin{equation}
\frac{F}{F^0} = \frac{\gamma_{exc}(\lambda_{exc})}{\gamma_{exc}^{0}(\lambda_{exc})}\cdot\frac{q(\lambda_{em})}{q^{0}(\lambda_{em})}\cdot\frac{\eta(\lambda_{em})}{\eta^{0}(\lambda_{em})},
\label{eq:TheoryEF}
\end{equation}

\noindent
where $\gamma_{exc}(\lambda_{exc})$ is the excitation rate at the wavelength of excitation, $q(\lambda_{em})$ is the quantum efficiency at the emission wavelength and $\eta(\lambda_{em})$ is the collection efficiency at the emission wavelength. All terms with $^0$ denote the same values in the absence of nano-antenna resonant effects for a monolayer on a flat SiO$_2$ substrate. The first term can be described as an enhancement of the rate of excitation, which is a result of the absorption of incoming light in the monolayer forming excitons, which will later recombine and lead to photoluminescence. As a reference, the number of excitons created within the monolayer at the excitation powers used in our experiments (0.02 $\mu J/cm^2$) is less than $10^{10} cm^{-2}$ \cite{Mouri2014a}. As the excitation rate is proportional to the electric field intensity $(|E|/|E^{0}|)^2$, the ratio of the rates can be estimated by simulating the electric field intensity both on the nano-antenna as well as on the SiO$_2$ substrate. The second factor describes an enhancement of the efficiency of emission once the monolayer is excited. This is described by the number of excitons which recombine to emit light ($q^{0} = \frac{\gamma_{r}}{\gamma_{r}+\gamma_{nr}}$) where $\gamma_{r}$ is the radiative emission rate and $\gamma_{nr}$ is the non-radiative emission rate which may be due to exciton dissociation and charge transfer or auger recombination among others. When a Purcell factor is applied to this quantum efficiency, it becomes $q = \frac{F_{P}\gamma_{r}}{F_{P}\gamma_{r}+\gamma_{nr}}$. As the quantum efficiency of monolayer WSe$_2$ ranges from 0.06$\%$ to 5$\%$ \cite{Mohamed2017a,Kim2019,Roy2020} and therefore $\gamma_{nr}>>\gamma_{r}$, the enhancement of $q/q^{0}$ can simply be approximated as $F_P$. The last factor represents an enhancement of the percentage of light which is emitted in the direction of the collection optics ($\eta(\lambda_{em})$). The nano-antenna structures are designed to emit a larger portion of light towards the collection optics, however, their high refractive index will lead to emission towards the substrate, therefore the overall change in collection efficiency for monolayer emission from the nano-antenna region is negligible when compared to that on flat SiO$_2$. We have simulated the three factors contributing to the WSe$_2$ monolayer fluorescence on flat SiO$_2$ and on dimer nano-antennas with a height of 135 nm, a gap of 150 nm and range of radii (r = 100 - 250 nm) which includes those measured for the three exemplary antennas shown in Fig. 2 of the main text. 

\begin{figure}[H]
	\centering
  \includegraphics[width=0.983\linewidth]{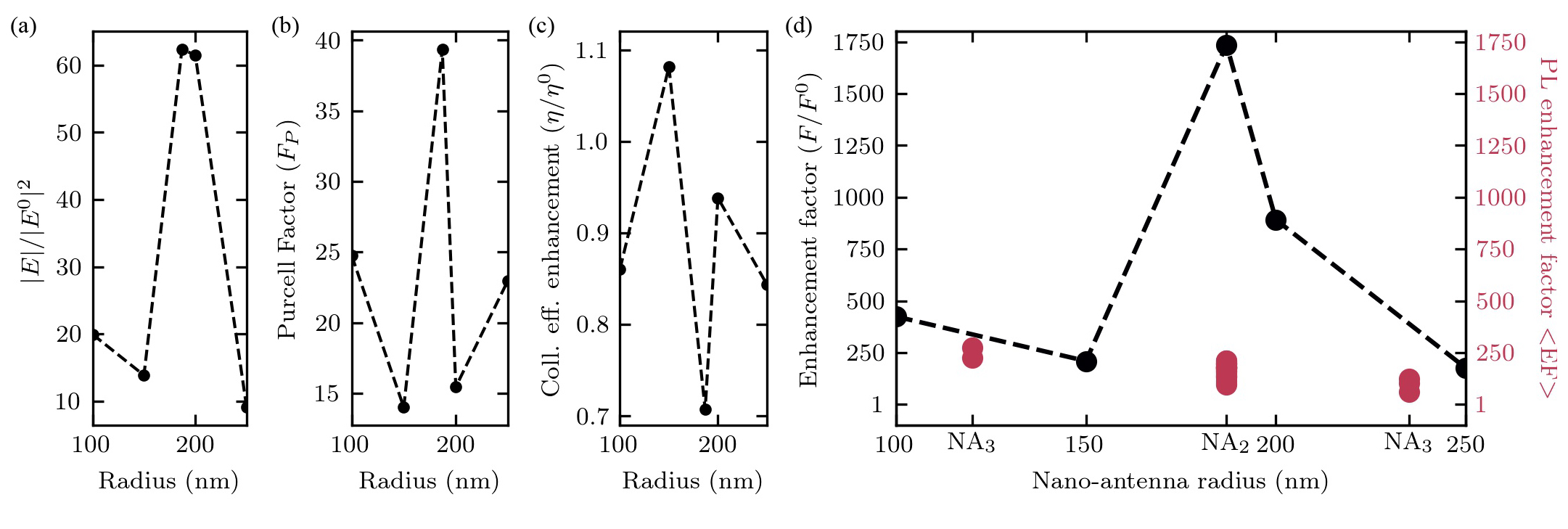}
  \caption{\textbf{Fluorescence enhancement simulations compared to experimental enhancement factors.} 
	\textbf{(a)} Simulation of the maximum electric field intensity hotspot 0.5 nm from the top inner vertex of the hexagonal dimer nano-antennas with a height of 135 nm, a gap of 150 nm and a range of radii (100-250 nm) compared to a position 0.5 nm from flat SiO$_2$. Wavelength = 638 nm. 
	\textbf{(b)} Simulation of the maximum Purcell enhancement factor for dipole emitter placed 0.5 nm from the top inner vertex of the dimer nano-antennas with the same geometry compared to a dipole placed 0.5 nm from a flat SiO$_2$ substrate. Wavelength = 765 nm.
	\textbf{(c)} Simulation of the collection efficiency enhancement for a dipole placed in the same positions as for the Purcell factor simulations. Percentage of light emitted within a cone defined by a numerical aperture (0.7) is compared to the same value simulated for a dipole 0.5 nm from a flat SiO$_2$ substrate. Wavelength = 765 nm. 
	\textbf{(d)} Calculation of the total fluorescence enhancement factor $F/F^0$ according to equation \ref{eq:TheoryEF} using the values simulated in \textbf{(a)}-\textbf{(c)} compared to experimental enhancement factors obtained from monolayer WSe$_2$ PL at nano-antenna sites.}
  \label{SI_PLEF}
\end{figure}

The maximum excitation rate enhancement, dependent on the electric field intensity, is displayed in Fig. \ref{SI_PLEF}(a) for a wavelength of 638 nm, which is identical to that of the pulsed laser used for the PL experiments. The excitation rate is highest for a dimer nano-antenna with a radius close to 200 nm. The maximum emission rate enhancement, dependent on the Purcell factor ($F_P$), is displayed in Fig. \ref{SI_PLEF}(b) for an emission wavelength of 765 nm, where the peak of the red-shifted PL emission from the WSe$_2$ monolayer on a majority of the nano-antennas was recorded. It is evident that the Purcell factor is highest for a radius close to the one measured for dimer NA$_2$. This suggests that the photonic modes are close to being in resonance with the monolayer emission for this size of dimer and further confirms that the emission rate enhancement can be tuned with nano-antenna size. Fig. \ref{SI_PLEF}(c) shows the collection efficiency enhancement that is simulated as a percentage of light emitted within the numerical aperture (0.7) of our experimental objective for a dipole emitter. All enhancements are calculated as ratios of the simulated values on the nano-antenna to those on flat SiO$_2$.

The maximum simulated fluorescence enhancement factor $F/F^0$ due to the nano-antenna resonances, which can contribute to the increased WSe$_2$ monolayer PL is plotted in Fig. \ref{SI_PLEF}(d) together with the experimental enhancement factors (red dots). The measured $\left\langle EF \right\rangle$ falls below the maximum simulated values due to the fact that monolayer emission is not only collected from the position of maximum enhancement, but also from surrounding areas which may not yield such high photonic enhancement. Another contributing factor to the low measured values is due to the fact that the monolayer may not conform to each dimer nano-antenna similarly and therefore may not couple to some nano-antenna hotspots as well as others. This is due to the monolayer transfer procedure, which is not controllable enough to guarantee a uniform coupling for all nano-antennas. What can be observed, however, is that the expected enhancement factor, mostly due to the excitation and emission rate enhancements, is expected to yield higher PL emission for a radius close to that of NA$_2$ as opposed to NA$_1$ or NA$_3$. This suggests that the coupling of the WSe$_2$ monolayer PL with the photonic resonance can be modulated by changing the nano-antenna size. 

\begin{center}
\Large
\textbf{Supplementary Note 6: Polarization dependent PL lifetimes of WSe$_2$ monolayer emission on dimer nano-antennas}
\end{center}

\begin{figure}[H]																																					
	\centering
  \includegraphics[width=0.983\linewidth]{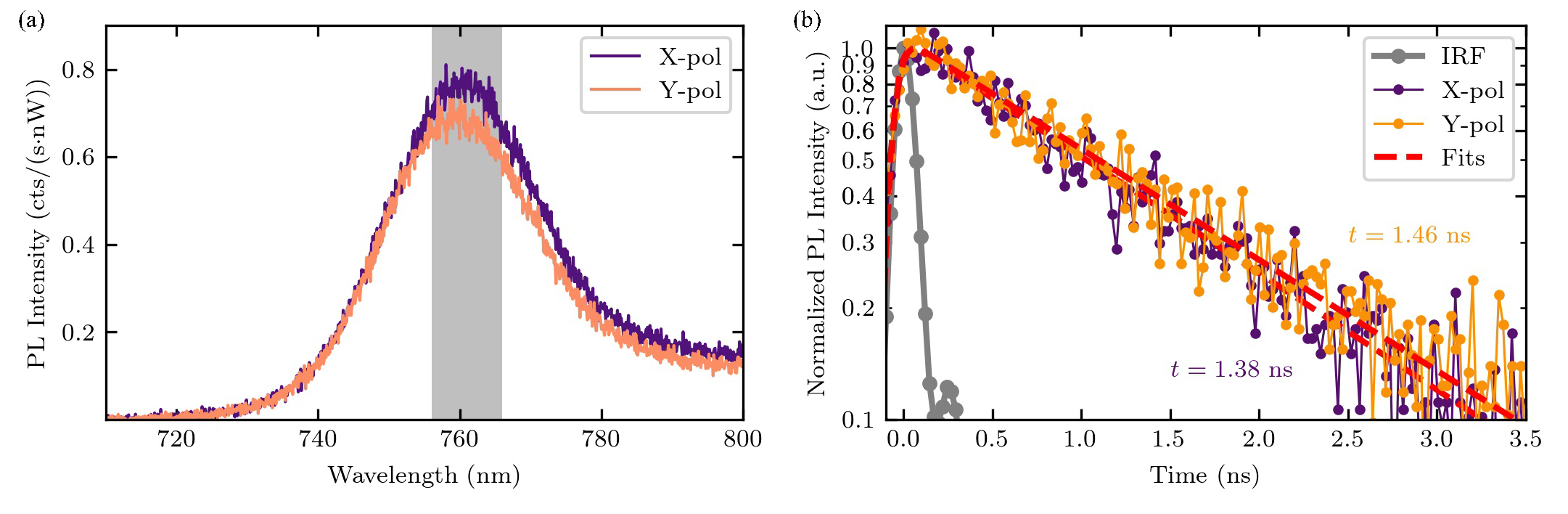}
  \caption{\textbf{Polarization dependent PL intensity and decay.} 
	\textbf{(a)} Room temperature monolayer WSe$_2$ PL spectra for an excitation (638 nm) polarization aligned parallel to the dimer axis (X-pol) and perpendicular to it (Y-pol) for a nano-antenna with a height of 135 nm, a gap of 150 nm and a radius of 190 nm. The PL intensity for the X-pol mode within the wavelength range in the gray area is 5$\%$ higher than for the Y-pol mode.
	\textbf{(b)} PL decay traces for the X-pol and Y-pol modes on the same dimer nano-antenna as shown in \textbf{(a)}. Red dashed lines are fits of the traces, which yield decay lifetimes of 1.38 ns and 1.46 ns. The X-pol mode coupled emission yields a 2.5$\%$ shorter decay time than that of the Y-pol mode. The gray trace represents the instrument response function.}
  \label{SI_G5}
\end{figure}

Monolayer WSe$_2$ room temperature PL emission was recorded over a range of dimer nano-antennas (11) for two perpendicular excitation polarizations corresponding to the X-pol and Y-pol modes. Fig. \ref{SI_G5}(a) displays the PL spectra recorded for an exemplary nano-antenna site with a height of 135 nm, a gap of 150 nm and a radius of 190 nm. We observed a higher emission intensity for the X-pol mode as expected from the higher electric field intensities for this polarization. We subsequently measured the emission lifetime of the WSe$_2$ monolayer PL in a 10 nm wavelength window centered at the peak (gray region in Fig. \ref{SI_G5}(a)) for both excitation polarizations, shown in Fig. \ref{SI_G5}(b). We discovered a small decay time shortening for the X-pol mode compared to the Y-pol mode, which was repeated for the majority of the measured dimer nano-antenna sites. This also suggests that Purcell enhancement is present in the emission of most of the measured structures and can be modulated by a rotation of the excitation polarization, coupling the PL emission to either the X-pol or Y-pol mode.

Subsequently, we integrated the PL intensity within the 10 nm wavelength shown as a gray region in Fig. \ref{SI_G5}(a) and calculated an increase of 5$\%$ in the PL intensity. The shortening of the lifetime seen in Fig. \ref{SI_G5}(b) was calculated to be 2.5$\%$. This correlates with the simulations of the fluorescence enhancement factor shown in Fig. \ref{SI_PLEF}, where Purcell enhancement contributed to approximately half of $F/F^0$. 

\begin{center}
\Large
\textbf{Supplementary Note 7: Second harmonic generation enhancement due to coupling with a dimer anapole mode}
\end{center}

\begin{figure}[H]
	\centering
  \includegraphics[width=0.747\linewidth]{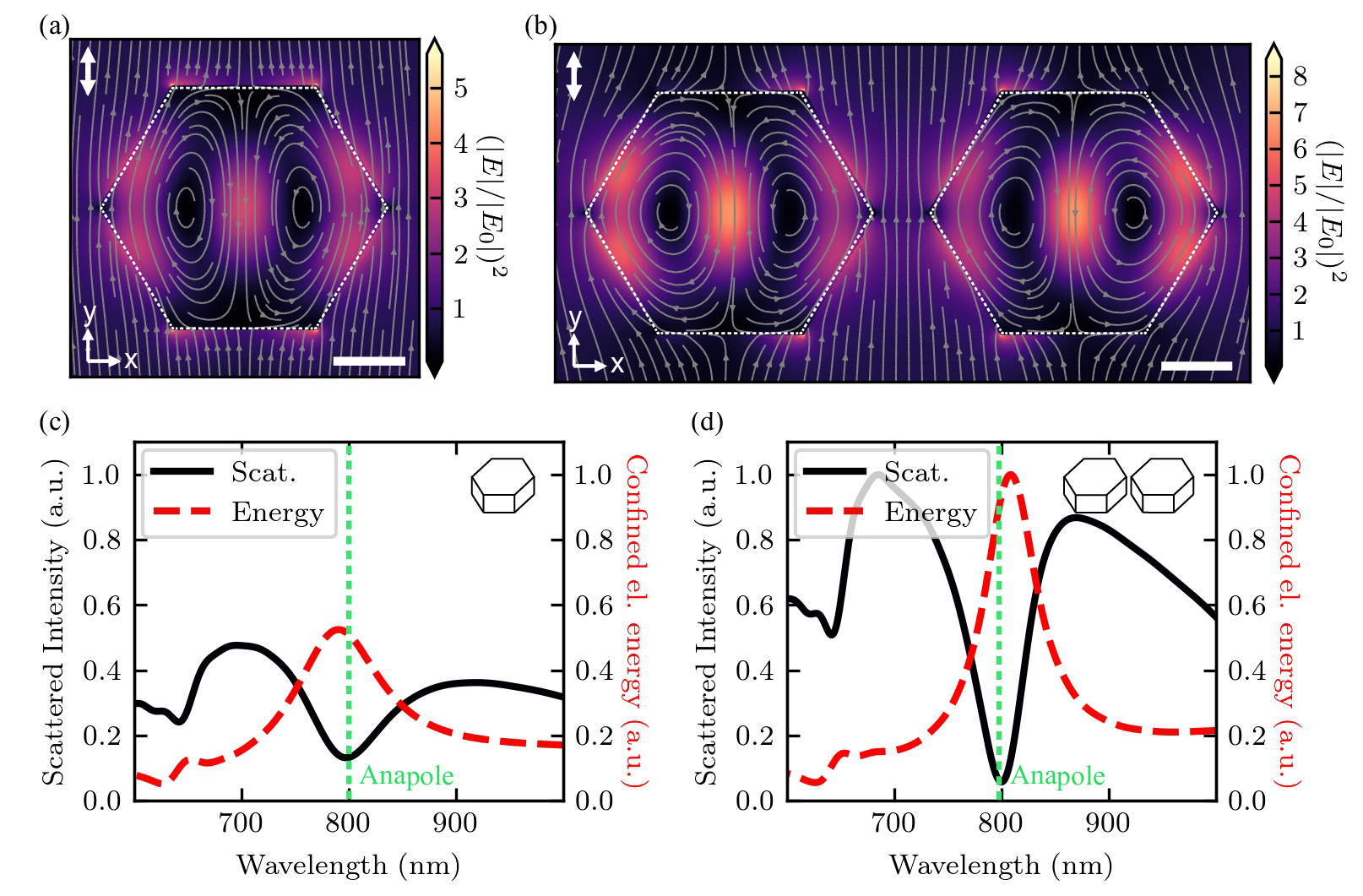}
  \caption{\textbf{Comparison of near-field electric field intensities and confined electric energies for the monomer and dimer anapole modes.}
	\textbf{(a),(b)} Electric field intensity spatial distributions for a cross-section at mid-height (30 nm) for a monomer (r = 200 nm, h = 60 nm) and dimer (r = 200 nm, h = 60 nm, g = 70 nm) nano-antenna respectively at the anapole mode wavelength. The double arrows in the top left of both \textbf{(a)} and \textbf{(b)} indicate the polarization direction of the excitation. The grey lines represent the electric field in and surrounding the nano-antenna at this cross section. Dashed white outlines represent the physical edges of the structure. Scale bars = 100 nm. 
	\textbf{(c),(d)} Simulated scattering cross section and confined electric energy inside the mid-height cross section of the monomer and dimer nano-antennas respectively. The scattering and electric energy spectra are normalized to the maximum of the dimer response for comparison. The vertical dashed, green line indicates the position of the anapole mode.}
	\label{MonoDimAnapoleComp}
\end{figure}

Dielectric nano-antenna anapole modes have previously been used for second harmonic generation enhancement \cite{Cambiasso2017,Busschaert2020,Tselikov2021}. This is done by coupling an excitation beam to an anapole mode. As the anapole mode increases the electric field intensity inside the material of the nano-antenna, this leads to a stronger SHG signal. In previous studies involving TMDs, this has been accomplished using the anapole mode of a circular monomer nano-antenna \cite{Busschaert2020,Tselikov2021}. As seen from Supplementary Note 1, the fabricated hexagonal WS$_2$ monomer nano-antennas also host anapole modes. Moreover, WS$_2$ hexagonal dimer nano-antennas have also been shown to host anapole resonances in Supplementary Note 3. Similar to dipole or quadrupole resonances, the anapole modes in monomers will hybridize when a dimer nano-antenna is formed. This hybridization will lead to confinement of a portion of the electric field intensity outside of the nano-antenna geometry. Figs. \ref{MonoDimAnapoleComp}(a) and (b) show the induced electric field intensity of the anapole mode in and surrounding a monomer (r = 200, h = 60 nm) and dimer (r = 200 nm, h = 60 nm, g = 70 nm) nano-antenna respectively at the midpoint of their height (30 nm) for an excitation polarization perpendicular to the dimer axis. However, it is interesting to note that electric field intensity anti-nodes in the dimer nano-antenna at this small gap separation are higher than those for the monomer signifying more tightly confined light. 

Second harmonic generation is a second order process, which, to a dipole approximation, is a surface phenomenon, which scales with the electric energy at the excitation wavelength squared, $|W_{E}^{(S)}|^{2}$ \cite{Cambiasso2017}. Therefore, we calculated the electric energy confined to the inside of the nano-pillars at the surface described for Figs. \ref{MonoDimAnapoleComp}(a) and (b) using the following definition \cite{Cambiasso2017}:

\begin{equation}
W_{E}^{(S)} = n^{2}/2\int\int\|E(r,\lambda_{A})\|^{2} dS,
\label{eqS1}
\end{equation}

\noindent
where $n$ is the refractive index and $\|E(r,\lambda)\|^{2}$ is the electric field intensity. The integration was carried out over the inside surface of the nano-antennas for a range of wavelengths. Figs. \ref{MonoDimAnapoleComp}(c) and (d) display the confined electric energy (red dashed curves) inside the monomer and the dimer nano-antenna respectively with a simulated scattering cross section (black solid curves) for each as a reference to the anapole mode. The scattering cross sections show a deeper minimum for the dimer anapole mode and an increased maximum confined electric energy (by a factor of 1.9) for the dimer nano-antenna when compared to the monomer. Two uncoupled monomers would yield a slightly higher electric energy than a dimer nano-antenna leading to higher SHG enhancement in the monomer when the electric energy is squared \cite{Cambiasso2017}.

\begin{figure}[H]
	\centering
  \includegraphics[width=\linewidth]{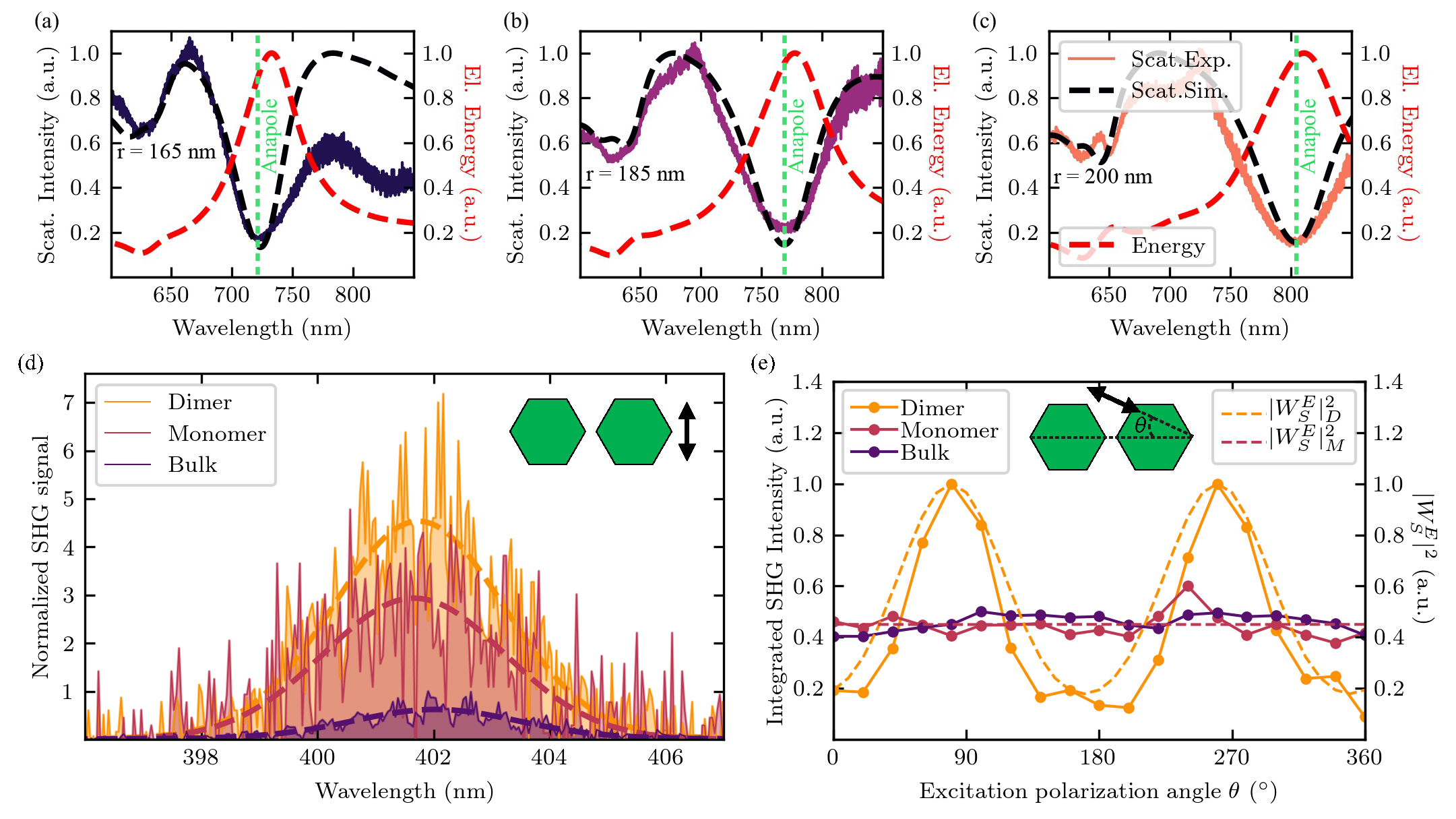}
  \caption{\textbf{Second harmonic generation enhancement of hexagonal dimer nano-antennas through coupling to an anapole mode.}
	\textbf{(a)-(c)} Dark field spectroscopy experiments (blue, purple, yellow curves) compared to scattering cross section simulations (black dashed curves) of dimer nano-antennas with different radii (r = 165 nm, r = 185 nm , r = 200 nm), a gap of 130 nm and a height of 60 nm. Dashed red curves correspond to simulated electric energy confined in the nano-antenna at mid-height. The anapole mode is highlighted with a vertical green dashed line.
	\textbf{(d)} Spectrum of second harmonic generation normalized to the excited area from a dimer nano-antenna (yellow) (r = 205 nm, h = 60 nm, g = 130 nm), a monomer nano-antenna (red) (r = 215 nm, h = 50 nm) and a bulk crystal (purple) (h = 60 nm). Excitation polarized perpendicular to the dimer axis as shown by the double black arrow in the inset. Dashed curves represent a Gaussian fit to the data.
	\textbf{(e)} Normalized, integrated polarization resolved second harmonic signal emitted from the dimer (yellow) (r = 200 nm, h = 60 nm, g = 130 nm) and monomer (red) (r = 240 nm, h = 50 nm) nano-antennas as well as the bulk crystal (purple) (h = 60 nm). The linear polarization of the excitation was set at an angle defined with respect to the dimer axis explained by the inset. Dashed curves represent the scaled polarization dependence of $|W_{S}^{E}|^{2}$ simulated for dimer (yellow) and monomer (red) nano-antennas with a similar geometry.}
  \label{FigSubSHG}
\end{figure}

In order to experimentally evaluate the SHG enhancement of the fabricated WS$_2$ hexagonal nano-antennas, we firstly identified anapole resonances in several dimers with a varying radius (r = 165 nm, r = 185 nm, r = 200 nm), a gap of 130 nm and a height of 60 nm by studying their scattering spectra. The scattering intensities of these structures, shown in Fig. \ref{FigSubSHG}(a)-(c) are in close agreement with FDTD simulations of the scattering cross sections for a similar geometry (black dashed curves). The presence of anapole resonances is corroborated by simulations of the electric energy inside each nano-antenna (red dashed curve) at the midpoint of its height. An indication that these resonances are anapole modes is the coincidence of a maximum in the confined electric energy and a minimum in the scattering cross section. As the anapole mode arises due to the destructive interference between an electric dipole and a magnetic toroidal resonance, it will not scatter light but instead confine it and increase the energy inside the dielectric structure \cite{Miroshnichenko2015}.

We subsequently recorded second harmonic generation spectra from a dimer (r = 205 nm, h = 60 nm, g = 130 nm) and monomer nano-antenna (r = 215 nm, h = 50 nm), which both host anapole modes near 800 nm, as well as from a bulk crystal (h = 60 nm). We used a 804 nm pulsed Spectra-Physics Ti-sapphire mode-locked fs laser excitation linearly polarized perpendicular to the dimer axis (Y-pol) with an average power of 1 mW. Fig. \ref{FigSubSHG}(d) displays the SHG spectra (collected via a Princeton Instruments spectrometer and CCD) normalized to the excited area. For the bulk measurement, the intensity was normalized to the area of the laser spot size ($A = \pi r^{2}$ for r = 700 nm). The hexagonal nano-antenna measurements were normalized to their respective areas ($A_{mono}=3r^{2}\sqrt{3}/2$ for the monomer with r = 215 nm and $A_{di}=3r^{2}\sqrt{3}$ for the dimer with r = 205 nm). We extracted an SHG enhancement factor by integrating the normalized second harmonic signal at the dimer nano-antenna position (yellow in Fig. \ref{FigSubSHG}(d)) and dividing by the similarly integrated signal from the bulk flake (purple dots), yielding a factor of 7.2. The normalized signal from the monomer was comparable to the dimer nano-antenna, which is expected from the simulations of the electric energy stored in the anapole mode of a similar geometry shown in Fig. \ref{MonoDimAnapoleComp}. 

We further studied the polarization dependence of the nano-antennas by shifting the excitation laser to 850 nm (3 mW), rotating its linear polarization and collecting integrated second harmonic signal through the use of an avalanche photo-diode (APD, ID Quantique). The laser wavelength used for these experiments was chosen due to the larger difference of confined electric energy expected for the X-pol and Y-pol modes of the dimer. The normalized polarization dependent SHG intensity was recorded for a dimer (r = 200 nm, h = 60 nm, g = 130 nm), monomer (r = 240, h = 50 nm) and bulk as shown in Fig. \ref{FigSubSHG}(e). Upon comparison of the measured signals, we observed a clear polarization of the second harmonic signal for the dimer nano-antenna while the same is not evident for the bulk or monomer nano-antenna measurements. As the excitation polarization is rotated, the X-pol and subsequently the Y-pol modes confine incident light differently leading to the observed polarization dependence. The electric field is mainly confined to the outside edges of the nano-antenna for the X-pol anapole mode, thereby not contributing to the SHG enhancement. For the Y-pol mode, however, the confinement is mostly inside the nano-antenna structure. The position of the confined electric field in the monomer and bulk does not change, therefore, leading to no polarization dependence of the SHG enhancement. We compared the experimental data to simulations of the relative SHG signal (calculated as the square of the electric energy as described above) with a varying linear polarization, shown in Fig. \ref{FigSubSHG}(e) as dashed curves, and observed good agreement for both structures.

\begin{figure}[H]
	\centering
  \includegraphics[width=\linewidth]{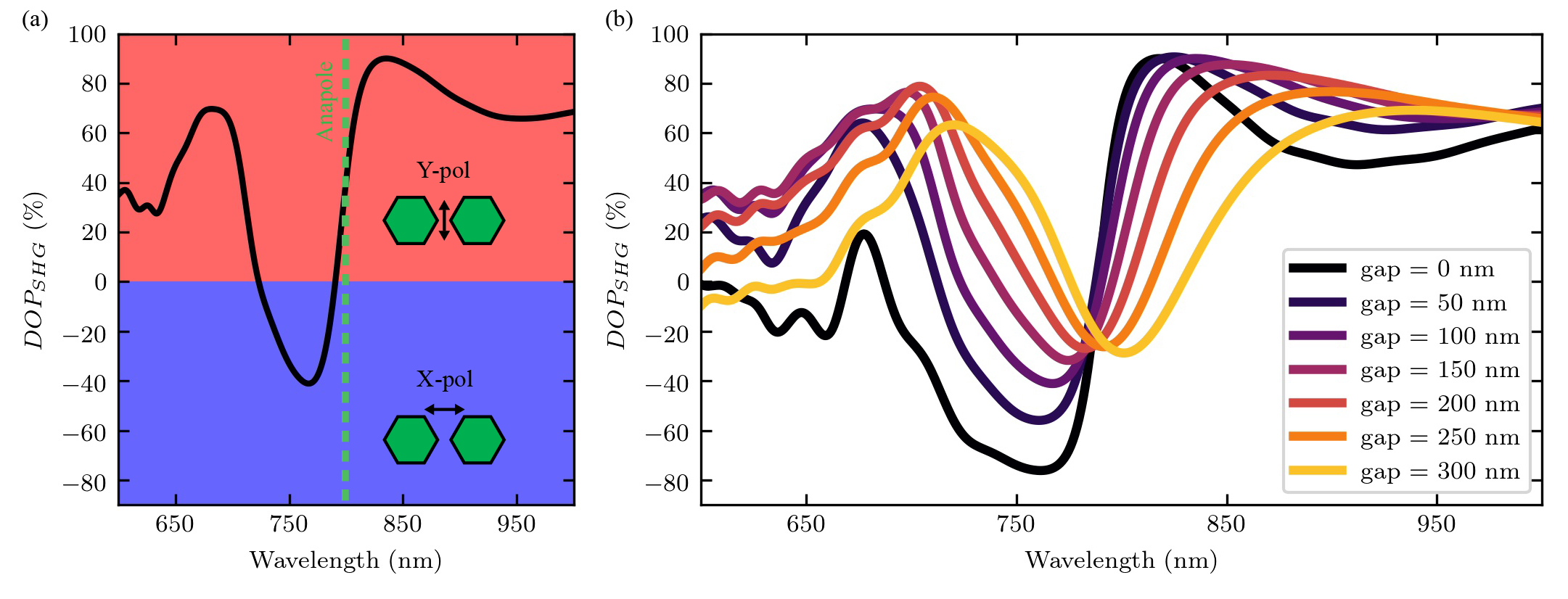}
  \caption{\textbf{Study of the degree of linear polarization of the SHG enhancement due to the anapole resonance in WS$_2$ dimer nano-antennas.}
	\textbf{(a)} Degree of linear polarization of the SHG enhancement over a range of excitation wavelengths for a WS$_2$ dimer nano-antenna with h = 60 nm, r = 200 nm, g = 100 nm. The position of the anapole resonance, at 800 nm, is shown by a green dashed line. The red (blue) region corresponds to a higher (lower) SHG enhancement in the Y-pol as compared to the X-pol mode shown by the insets.
	\textbf{(b)} Gap dependent degree of linear polarization of the SHG enhancement over a range of excitation wavelengths.  Smaller dimer gaps lead to higher degrees of linear polarizations.}
  \label{SHG_DOP}
\end{figure}

Additionally, we studied the degree of linear polarization of the SHG enhancement that we discovered in the WS$_2$ dimer nano-antennas. We performed simulations of the square of the electric energy inside the nano-antenna structure at the anapole resonance $|W_{E}^{S}|^2$ for the X-pol and Y-pol modes and defined a degree of linear polarization as follows:

\begin{equation}
DOP_{SHG} = \frac{|W_{E}^{S}|_{Y}^{2}-|W_{E}^{S}|_{X}^{2}}{|W_{E}^{S}|_{Y}^{2}+|W_{E}^{S}|_{X}^{2}},
\label{eqS2}
\end{equation}

\noindent
where $|W_{E}^{S}|_{X}^{2}$ and $|W_{E}^{S}|_{Y}^{2}$ represent the square of the electric energy confined in the anapole mode for the X-pol and Y-pol modes respectively. Fig. \ref{SHG_DOP}(a) shows the $DOP_{SHG}$ over a range of excitation wavelengths for a dimer nano-antenna of 60 nm height, 200 nm radius, 100 nm gap and an anapole resonance at 800 nm. The highest $DOP_{SHG}$ for this geometry is >80$\%$. An important observation is that degree of linear polarization is positive for some excitation wavelengths and negative for others, indicating that the mode, which will provide the highest SHG enhancement at that wavelength is the Y-pol and X-pol mode respectively. This suggests that by varying the excitation wavelength, one can induce opposite SHG enhancement polarizations in the dimer nano-antenna and therefore, modify the polarization of the SHG signal. This can be advantageous for the realization of optical logic gates as proposed in reference \cite{Bovino2009}. 

We subsequently studied the change in the degree of polarization over a range of gap sizes. Fig. \ref{SHG_DOP}(b) displays the excitation wavelength dependencies for gaps of 0-300 nm. As the gap increases, the degree of linear polarization of the SHG enhancement is reduced over the entire wavelength range. This is expected as the increase in gap leads to a dimer nano-antenna with less hybridized resonances and therefore a response closer to that of the monomer nano-antenna, which we have shown to exhibit no polarization dependence. Incidentally, the highest degrees of linear polarization of the SHG enhancement due to both the X-pol and Y-pol anapole resonances are observed for gaps below 50 nm, which can only be achieved through the AFM repositioning technique introduced in the main text. This highlights the importance of WS$_2$ dimer nano-antennas with ultra-small gaps for non-linear light applications.

\begin{center}
\Large
\textbf{Supplementary Note 8: Further Atomic force microscopy repositioning experiments}
\end{center}

\begin{figure}[H]
	\centering
  \includegraphics[width=0.903\linewidth]{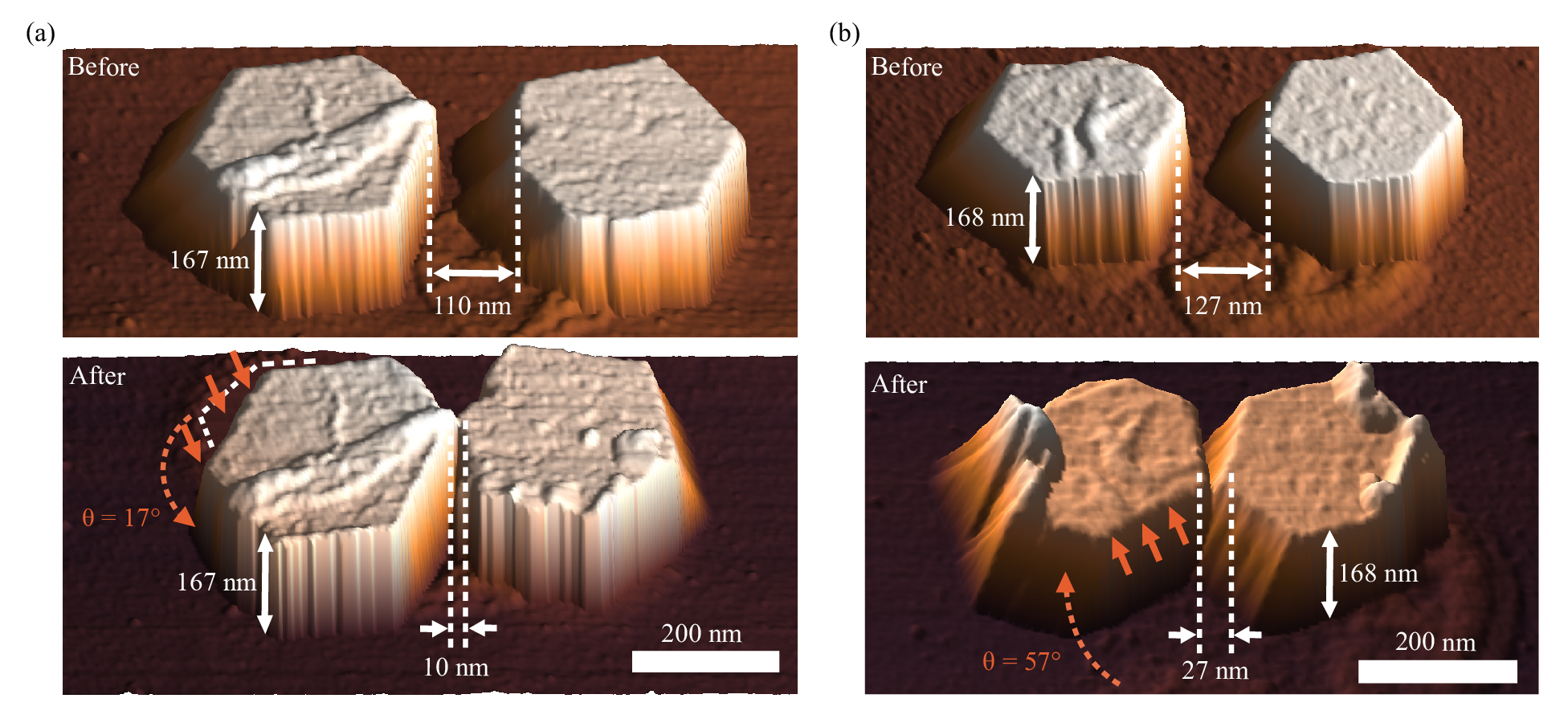}
  \caption{\textbf{Additional exemplary AFM scans before and after repositioning manipulations.} 
	\textbf{(a)} AFM scans of a dimer nano-antenna before and after repositioning with a radius of 163 nm, a height of 167 nm, a reduction in gap from 110 nm to 10 nm and a rotation of 17$^{\circ}$.
	\textbf{(b)} AFM scans of another dimer nano-antenna before and after repositioning with a radius of 140 nm, a height of 168 nm, a reduction in gap from 127 nm to 27 nm and a rotation of 57$^{\circ}$.}
  \label{AFMSI}
\end{figure}

In order to provide evidence for the reproducibility of the atomic force microscope (AFM) dimer repositioning method, we show before and after scans (Fig. \ref{AFMSI}) of two more nano-antennas which have been repositioned. In Fig. \ref{AFMSI}(a) we achieve a rotation of 17$^{\circ}$ as well as a reduction of the dimer gap from 110 nm to 10 nm, which is the smallest achieved in our experiments, for a dimer with a radius of 163 nm and a height of 167 nm. In Fig. \ref{AFMSI}(b) we show a rotation of 57$^{\circ}$ and a gap reduction from 127 nm to 27 nm for a dimer with a radius of 140 nm and a height of 168 nm.

\begin{center}
\Large
\textbf{Supplementary Note 9: Radius of curvature measurement}
\end{center}

\begin{figure}[H]
	\centering
  \includegraphics[width=0.967\linewidth]{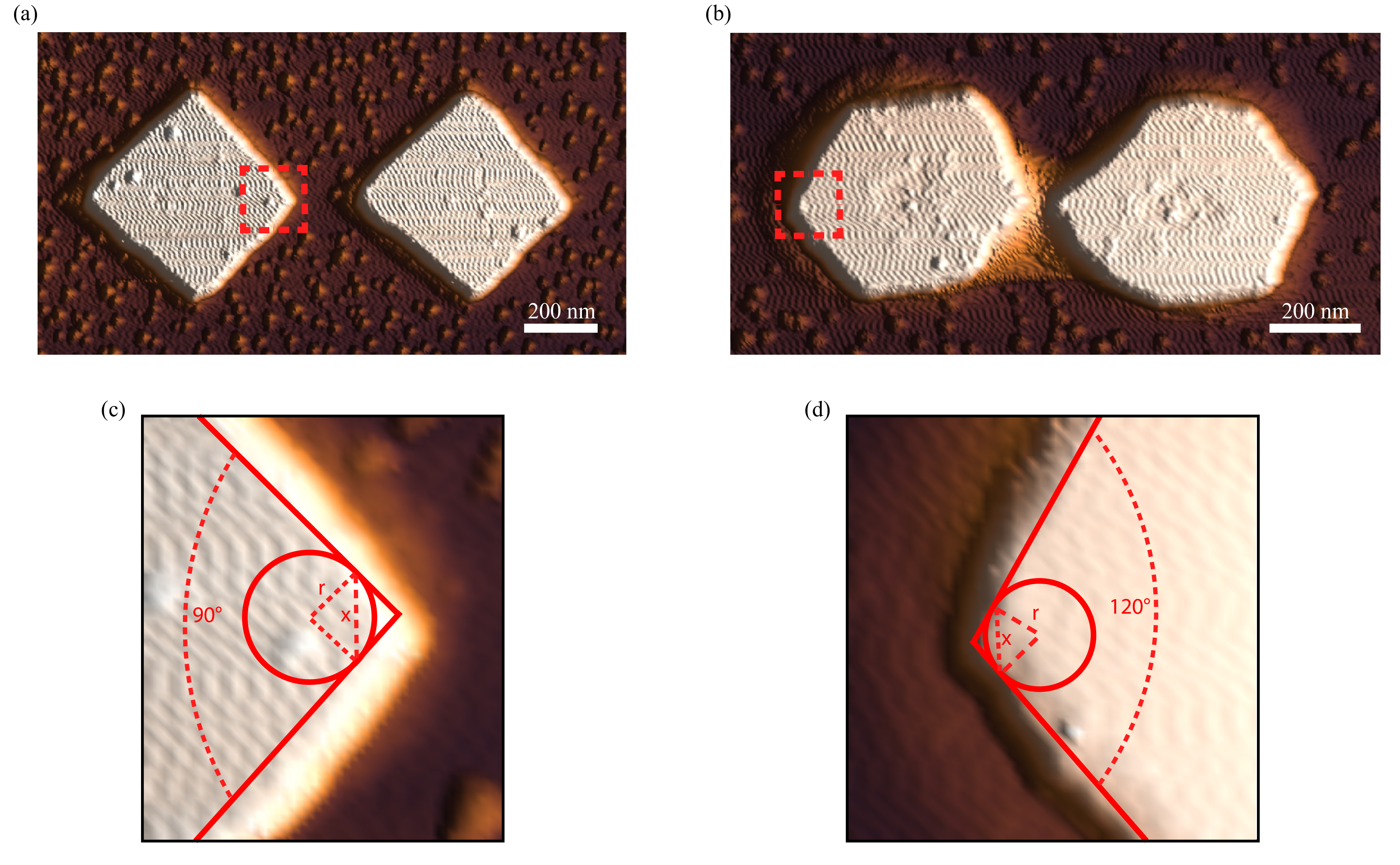}
  \caption{\textbf{AFM scans showing method of extracting vertex radius of curvature.}
	\textbf{(a),(b)} AFM scans of exemplary square (r = 265 nm, h = 84 nm, gap = 220 nm) and hexagonal (r = 240 nm, h = 135 nm, gap = 90 nm) dimer nano-antennas.
	\textbf{(c),(d)} Portions of the previous AFM scans shown as dashed red squares in \textbf{(a)} and \textbf{(b)}. These are overlapped with solid red lines at 90$^{\circ}$ and 120$^{\circ}$ to each other showing the top edges of the structures with an inscribed solid circle, which is used to extract the radius (r) of curvature of the vertex. Distance measured from the AFM scan is denoted by x, which is the shortest distance between the top two edges of the structure. }
  \label{RadOfC}
\end{figure}

In order to provide simulations of the electric field and Purcell enhancement for realistic structures, we measured the radius of curvature of vertices of many of our fabricated structures with atomic force microscopy and used recorded values in further simulations including those seen in Fig. 4 of the main text and in Supplementary Note 10. Since many of the square and hexagonal structures exhibited a radius of curvature below or on the order of the maximum resolution of our AFM measurements, we used a geometric approach described below to reliably measure the radius of curvature.

In Figs. \ref{RadOfC}(a) and (b), we have shown an example of the square (r = 265 nm, h = 84 nm, gap = 220 nm) and hexagonal (r = 240 nm, h = 135 nm, gap = 90 nm) dimer nano-antenna AFM scans from which we extracted values of the radius of curvature of the vertices. For both geometries, we collected data from the top surface of the nano-antennas where the bottom of the AFM tip made contact with the structure to provide the highest resolution possible. The extracted value from the AFM scans denoted as x in Figs. \ref{RadOfC}(c) and (d) is the shortest distance between two edges which are at 90$^{\circ}$ or 120$^{\circ}$ relative to each other for the square and hexagonal geometries respectively. This can be thought of as the base of a triangle formed by two radii of curvature that perpendicularly intersect the edges of the respective structure and the angle between them can easily be calculated to be 90$^{\circ}$ and 60$^{\circ}$ for the square and hexagonal geometries respectively. The radius denoted as r in the figure can then be computed using the Pythagorean theorem.

\begin{center}
\Large
\textbf{Supplementary Note 10: Modulation of the electric field intensity and Purcell factor of WS$_2$ dimer nano-antennas}
\end{center}

\begin{figure}[H]
	\centering
  \includegraphics[width=0.983\linewidth]{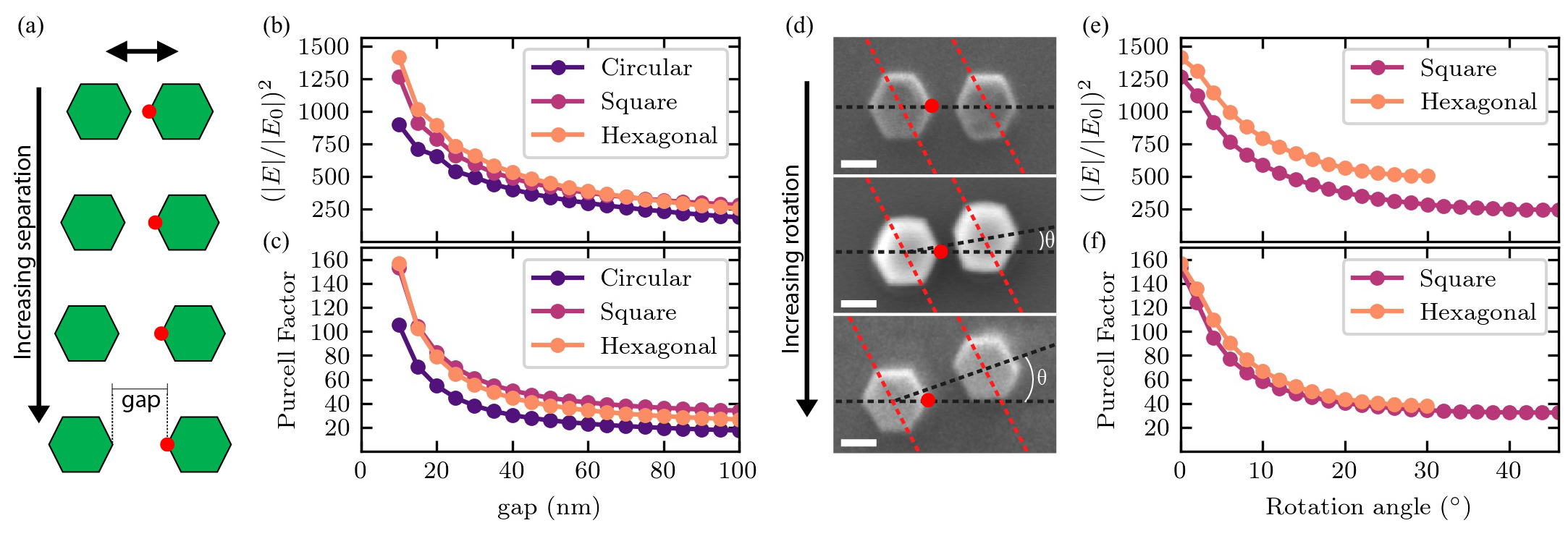}
  \caption{\textbf{Modulation of the electric field intensity and Purcell factor in dimer nano-antennas.} 
	\textbf{(a)} Illustration of the change in dimer gap used for the simulations in \textbf{(b)} and \textbf{(c)}. 
	\textbf{(b),(c)} Gap dependent maximum electric field intensity and Purcell factor within the hotspot of the optimized dimer designs for each geometry at the previously used wavelengths.
	\textbf{(d)} Top-view SEM images of hexagonal dimer nano-antennas demonstrating the ability to utilize the crystal symmetry to fabricate structures with a rotation ($\theta$) relative to the dimer axis. Red circles indicate position of electric field hotspot or dipole source for Purcell enhancement in \textbf{(e)} and \textbf{(f)}. Scale bars = 200 nm.
	\textbf{(e),(f)} Rotation angle dependent maximum electric field intensity and Purcell factor within the hotspot of the optimized dimer designs for the square and hexagonal geometries at the previously used wavelengths.}
  \label{SI_E_P_Modulate}
\end{figure}

There are several methods of modulating the electric field confinement and the local density of optical states governing the Purcell effect within the hotspots of the dimer nano-antenna resonances. One of these methods is to increase the gap separation of the dimer constituent nano-pillars. This approach can either be achieved during the patterning stage of the fabrication process or, to a greater precision, using the AFM repositioning method. We numerically studied the effect of moving the two nano-pillars apart from each other (schematically depicted in Fig. \ref{SI_E_P_Modulate}(a)) on the electric field intensity and Purcell factor at the hotspots in Fig. \ref{SI_E_P_Modulate}(d) and (e). We also compared these effects for the three dimer designs. Both the electric field intensity and Purcell factor decrease exponentially with separation distance. The hexagonal structure yields the highest electric field confinement at small separations, however the intensity induced by the square structure is highest for large gap values (>75 nm). The Purcell factor of the square structure, however, exceeds that of the hexagonal at separation values as low as 20 nm.

Another approach available only to the hexagonal and square nano-antennas takes into account the control of the alignment of the two nano-pillars relative to each other. One method to achieve such alignment was previously shown in Fig. 3 of the main text, while another method utilizes the chemical etching mechanism, which etches faster in the armchair axis of the crystal. Therefore, two nano-pillars fabricated spatially close to each other will always have the same orientation with respect to the symmetry of the original crystal. However, if the dimer nano-antenna design is rotated at the electron beam patterning stage of the fabrication procedure, as displayed in Fig.\ref{SI_E_P_Modulate}(d), the relative orientation of the antennas will have changed along the new axis connecting their centers, thereby, positioning their vertices closer or further apart from each other. As shown in Fig.\ref{SI_E_P_Modulate}(e) and (f), this can lead to a more subtle modulation of the electric field intensity and the Purcell enhancement for a quantum emitter located within the hotspots. The range of angles available to the hexagonal geometry is limited to 30$^{\circ}$,  while the square structure can be rotated to 45$^{\circ}$ before the electric field intensity and Purcell factor begin to increase once more. The large modulation of both electric field intensity and Purcell factor shown here requires a small gap separation. As the gap increases, the rotation will lead to smaller changes in the electric field intensity and Purcell factor.

\newpage
\bibliographystyle{unsrt}
\bibliography{./library}